\documentclass[10pt]{article}

\usepackage{mathptmx} 
\usepackage[T1]{fontenc}  
\usepackage[utf8]{inputenc}
\usepackage{lipsum}  
\usepackage{eqparbox}
\usepackage{amsmath, amssymb, graphics, setspace}
\RequirePackage[english]{babel}
\RequirePackage{blindtext}
\RequirePackage{lipsum}
\usepackage{romannum}
\usepackage{soul}
\usepackage{scrextend}
\usepackage{amsmath}
\RequirePackage{graphicx}
\makeatletter
\ifx\pdfoutput\undefined\else
\fi
\makeatother

\RequirePackage{float} 
\RequirePackage{multirow}
\RequirePackage{makecell}
\RequirePackage{booktabs}
\usepackage{fancyhdr}
\usepackage{float} 

\usepackage{graphicx, array, booktabs, adjustbox}

\usepackage{xcolor}
\RequirePackage{caption}
\usepackage{bigfoot}
\usepackage[hyperfootnotes=false]{hyperref}
\hypersetup{
  colorlinks=false,
  pdfborder={0 0 0}
}
\usepackage[a4paper, margin=1in]{geometry}
\usepackage{authblk}
\usepackage{caption} 
\makeatletter
\def\blfootnote{\xdef\@thefnmark{}\@footnotetext}
\makeatother

\newenvironment{keywords}{
    \noindent\textbf{Keywords: }\itshape
}{}

\begin{document}

\title{\textbf{\Large {XVertNet: Unsupervised Contrast Enhancement \\
of Vertebral Structures with Dynamic \\ 
Self-Tuning Guidance and Multi-Stage Analysis}}}

\author{\normalsize{Ella Eidlin\textsuperscript{1}, Assaf Hoogi\textsuperscript{2}, Hila Rozen\textsuperscript{3}, Mohammad Badarne\textsuperscript{4}, and Nathan S. Netanyahu\textsuperscript{5}}}

\date{}

\pagenumbering{arabic}
\maketitle

\begin{samepage}
\begin{flushleft}

\blfootnote{
\hspace{-15pt}\textsuperscript{1} Department of Computer Science, Bar-Ilan University, Ramat Gan, Israel (e-mail: ella.0519@gmail.com). \\
\textsuperscript{2} School of Computer Science and the Data Science and Artificial Intelligence Research Center, Ariel University, Ariel, Israel (e-mail: assafh@ariel.ac.il). \\
\textsuperscript{3} Department of Emergency Medicine, Hadassah Hebrew University Medical Center, Ein Kerem Campus, Jerusalem, Israel (e-mail: hila.rozen1@gmail.com). \\
\textsuperscript{4} Department of Radiology, Wolfson Medical Center, Holon, Israel (e-mail: mohammad.badarny@gmail.com). \\
\textsuperscript{5} Department of Computer Science and the Data Science and AI Institute, Bar-Ilan University, Ramat Gan, Israel (e-mail:nathan@cs.biu.ac.il); also affiliated with the Department of Computer Science, College of Law and Business, Ramat Gan, Israel.}
\end{flushleft}
\end{samepage}

\begin{abstract}

Chest X-ray is one of the main diagnostic tools in emergency medicine, yet its limited ability to capture fine anatomical details
can result in missed or delayed diagnoses. To address this, we introduce XVertNet, a novel deep-learning framework designed
to enhance vertebral structure visualization in X-ray images significantly. Our framework introduces two key innovations:
(1) an unsupervised learning architecture that eliminates reliance on manually labeled training data—a persistent bottleneck
in medical imaging, and (2) a dynamic self-tuned internal guidance mechanism featuring an adaptive feedback loop for
real-time image optimization. Extensive validation across four major public datasets revealed that XVertNet outperforms state-of-the-art enhancement methods, as demonstrated by improvements in evaluation measures such as entropy, the Tenengrad
criterion, LPC-SI, TMQI, and PIQE. Furthermore, clinical validation conducted by two board-certified clinicians confirmed
that the enhanced images enabled more sensitive examination of vertebral structural changes. The unsupervised nature of
XVertNet facilitates immediate clinical deployment without requiring additional training overhead. This innovation represents
a transformative advancement in emergency radiology, providing a scalable and time-efficient solution to enhance diagnostic
accuracy in high-pressure clinical environments.

\end{abstract}


\section{Introduction} \label{introduction}
Chest X-ray imaging remains a cornerstone of clinical diagnosis, valued for its speed, accessibility, and ability to detect a broad range of thoracic and bone pathologies. From pleural effusions and aortic dissections to bone fractures, osteopenia, and scoliosis, a single chest X-ray can reveal critical diagnostic information when interpreted by experienced radiologists ~\cite{Laserson2018}. However, the inherent low spatial resolution
and limited soft-tissue contrast of X-rays often leads to
missed findings, especially during early disease stages ~\cite{Ramani2013}.
CT and MRI are both powerful alternative diagnostic tools ~\cite{Yacob2005}, yet their use in emergency settings presents significant
challenges as well. While CT is faster and more accessible, it exposes patients to radiation and often requires
contrast material. Moreover, its lack of mobility makes
it less suitable for immediate assessments. On the other
hand, MRI is generally impractical due to its high cost
and lengthy scan times ~\cite{Ginde2008}. These constraints frequently
necessitate the use of other imaging methods that are better suited to urgent care situations. Consequently, chest X-ray
Extended author information available on the last page of the article
remains the first-line imaging tool in emergency departments.
While radiology experts can often navigate the low specificity of these images, non-experts—including ER clinicians
and trainees—frequently struggle to detect subtle vertebral
abnormalities. Alarmingly, studies report a 20.35\% discrepancy between initial and final radiological assessments in
ERs, with 7.48\% being clinically significant ~\cite{Mattsson2018}. Another
study found that 13.5\% of X-ray interpretation errors affected
clinical decisions, especially in pediatric cases and among
less experienced readers ~\cite{Wilson2022}. Improving the visibility of key
spinal structures—particularly the vertebrae—is therefore
critical for accurate, timely diagnosis in high-pressure environments. Existing contrast enhancement methods tend to
fall short in the following sense. Most rely on supervised
learning with high-contrast annotations or curated training
datasets, which are scarce in medical imaging. To overcome
these barriers, we propose XVertNet, a novel method designed
to enhance vertebral contrast in chest radiographs without
the need for external labels or high-resolution ground-truth
images.
The key contributions of our approach are as follows:
\begin{itemize}
\item

\textbf{Dynamic Self-Adapted Internal Guidance Layer}:

We introduce a novel self-adapting internal \textit{guidance layer} (GL) that replaces external modules and knowledge distillation. It directly shapes activations to boost information flow and feature quality.

\item
\textbf{Fully Unsupervised U-Net Training}:

We present the first fully unsupervised, iterative \textit{U-Net framework} for medical image contrast enhancement—a practical solution to the lack of labeled medical data.

\item
\textbf{Clinical Utility}: Our approach enables accurate diagnosis of vertebral structures from non-specific chest radiographs, addressing a significant clinical need.
\item
\textbf{Robust Evaluation}: We validate our method across diverse public datasets with varying image characteristics (e.g., 8-bit and 12-bit pixel formats), demonstrating its effectiveness in handling high spatial diversity.
\end{itemize}

The rest of the paper is organized as follows. Section~\ref{Related Work} reviews existing works and their limitations in medical image contrast enhancement. 
Section~\ref{proposedmethod} outlines the design and training of XVertNet. Section~\ref{experimentsanalysis} presents extensive quantitative and qualitative evaluations, as well as an ablation study. Section~\ref{discussion} provides a discussion and conclusion. \\

\section{Related Work}
\label{Related Work}
Various techniques have been developed to enhance image contrast.

\textbf{Traditional methods} include \textit{contrast limited adaptive histogram equalization} (CLAHE) \cite{Pisano1998}, \textit{gamma correction} \cite{Ikhsan2014}, \textit{histogram equalization} (HE) \cite{Sundarama2011}, and \textit{edge-aware filters} \cite{He2016}. However, combining HE or CLAHE with other enhancement techniques often results in over-enhancement and loss of critical image details \cite{Ali2015, Wong2016, Singh2016, Wen2016, Qiu2017, Madmad2019, Siracusano2023, Avci2023}. These methods struggle to balance enhancement with the preservation of clinically relevant features, particularly when subtle tissue differences are essential for diagnosis. Notable approaches include Farbman \textit{et al.}'s \textit{weighted least squares} (WLS) filtering \cite{Fattal2008} for multi-scale detail manipulation, which relies on clear edges and textures. Ameen's swift algorithm \cite{Ameen2018} offers low computational complexity through a three-step enhancement process but tends to degrade subtle textures. Similarly, Lin \textit{et al.}'s gradient-domain guided image filtering \cite{LLi2022} integrates adaptive amplification but faces issues with computational complexity and parameter sensitivity.

\textbf{Learning-based approaches} have emerged as powerful alternatives. Supervised techniques \cite{Lore2017, Hu2017, Xiao2019, Rawat2021, Zhong2023, Li2023} effectively enhance fine details but require extensive labeled training data. In medical contexts, obtaining such data is challenging due to the high cost and time required for expert annotations. The scarcity of annotated medical datasets, combined with privacy concerns and regulatory requirements, further complicates data collection for supervised learning. This has spurred the development of unsupervised methods that do not rely on manual annotations.

Unsupervised approaches primarily fall into two categories, i.e., unpaired data and synthetic data. Unpaired data methods use independent collections of enhanced and non-enhanced images, employing extensions based on the \textit{generative adversarial network} (GAN) \cite{goodfellow2014}. These methods include \textit{CycleGAN} \cite{Zhu2017} and its variant \textit{Cycle-MedGAN} \cite{Armanious2019}, which introduces non-adversarial cycle losses, and the \textit{structured illumination constrained GAN} (StillGAN) \cite{Ma2021}, which implements illumination regularization. He \textit{et al.} \cite{He2023} proposed a content-aware loss for medical fundus image enhancement. Another notable approach is \textit{Zero-DCE} \cite{guo2020}, a zero-reference low-light enhancement method that estimates pixel-wise curves directly from the input using a self-supervised learning scheme. Zero-DCE avoids reliance on paired data or GAN training and instead optimizes multiple perceptual constraints, such as exposure and color consistency. 
Overall, these methods often struggle to preserve fine structures and rely on domain-specific features, which may not translate well across different imaging modalities—especially for X-rays, where the relevant context differs significantly from fundus images.

\textbf{Synthetic-based approaches} have also shown significant progress. Previous works by Gozes and Greenspan \cite{Gozes2018, Gozes2020} employed digitally reconstructed radiographs from CT images. More recent methods explore diverse synthesis techniques, such as \textit{contrast synthesis} in MRI \cite{Bone2021, Xue2022} and GAN-based \textit{CT synthesis} \cite{Pang2023, Franzes2023}.

The \textit{clinical oriented fundus correction network} (Cofe-Net) \cite{Shen2020} employs a generative model to synthesize enhanced retinal fundus images from low-quality inputs. By modeling the degradation process and incorporating a high-quality image prior, this framework learns to map low-quality images to high-quality counterparts. The approach ensures that synthesized images retain key anatomical structures while improving contrast and sharpness, facilitating better diagnostic accuracy. Also, the \textit{annotation-free restoration network for cataractous
fundus images} (ArcNet) \cite{Li2022} generates high-quality fundus images by learning from unlabeled data, using a restoration network that synthesizes clean images from cataract-degraded inputs. This method does not require manual annotations, making it a scalable and efficient solution for handling large datasets of degraded images.

Madmad and Vleeschouwer~\cite{Madmad2019} proposed a learning-based approach for X-ray contrast enhancement, using a synthetic dataset simulating background objects (lungs) and localized structures (nodules, fractures, and lung pipes). However, the model's generalizability is limited by the constrained diversity of the synthetic training data, reducing its effectiveness in broader clinical applications. Additionally, generating high-quality synthetic data that accurately reflects real-world scenarios is challenging due to complexities in medical imaging physics, such as scatter radiation, beam hardening, and artifacts. The \textit{source-free unsupervised domain adaptive medical image enhancement} (SAME) model \cite{Lin12023} offers a teacher-student approach but requires high-contrast images as a ground truth.

\textbf{Super-resolution techniques} provide another alternative for image enhancement. The \textit{zero-shot super\linebreak-resolution} (ZSSR) model  \cite{Shocher2018} leverages internal image statistics without external training, while a \textit{Cycle-in-Cycle GAN} (CinC-GAN)-based method \cite{cincgan2018} combines cycle-consistency with GAN-based super-resolution. The \textit{denoising super-resolution via variational autoencoder} (dSRVAE) model \cite{Liu2020} employs perceptual loss, which may introduce diagnostically problematic artifacts. Sander \textit{et al.} \cite{Sander2021} proposed an unsupervised autoencoder approach for enhancing anisotropic 3D cardiac MRI. Recent advancements in this area focus on improving architectural designs to handle the unique challenges of medical imaging, such as preserving diagnostic features while suppressing noise and artifacts. Despite these advances, achieving real-time performance with high-quality enhancement remains difficult.

\textbf{Medical image denoising methods}
have been developed for medical images. Classical methods like the \textit{block-matching and 3D} (BM3D) filtering ~\cite{sheng2014} and \textit{total variation} (TV) denoising have been widely adopted for their simplicity and effectiveness, but often compromised fine structural details. More recently, \textit{deep learning} (DL)-based approaches such as \textit{denoising convolutional neural networks} (DnCNNs) ~\cite{zhang2016}, \textit{Noise2Void} (N2V)~\cite{krull2019}, \textit{residual encoder-decoder convolutional neural network} (RED-CNN)~\cite{chen2017}, \textit{Deformed2Self}~\cite{jxu2021}, and \textit{edge enhancement based transformer} (Eformer)~\cite{luthra2021} introduced data-driven frameworks that leverage residual learning, perceptual loss, and attention mechanisms for denoising. These models significantly improve quality in low-dose CT and MRI applications. However, many rely on clean target data or generalize poorly to highly noisy or low-contrast regions, such as vertebral structures in chest X-rays. Our work addresses this gap by proposing a fully unsupervised enhancement pipeline focused on contrast enhancement of challenging regions such as the spine, where conventional denoising often fails to retain diagnostic detail.

Moreover, while both denoising and contrast enhancement seek to improve image quality, they approach the task from fundamentally different perspectives. Denoising techniques are designed to suppress random or structured noise, but do not specifically enhance contrast visibility or the clarity of structural details. In contrast, our work is exclusively focused on enhancing the perceptual contrast of anatomical structures, without incorporating or relying on any noise modeling. Thus, our comparative evaluation was confined to unsupervised contrast enhancement methods.

\textbf{Auxiliary guidance techniques} have recently been integrated to improve performance. The \textit{Laplacian pyramid super-resolution network} (LapSRN) \cite{Lai2017} and the \textit{structure-preserving super-resolution} (SPSR) \cite{Ma2020} methods use subbranches for multi-scale enhancement. The \textit{structure-consistent restoration network} (SCR-Net) \cite{HengLi2022} and the \textit{generic fundus image enhancement network} (GFE-Net) \cite{Liu2023} employ dual decoders for guided reconstruction, while \textit{ESDiff} \cite{FLiu2023} combines enhancement with vessel segmentation. However, these approaches often introduce computational overhead and potentially irrelevant features, as their guidance components rely on external training that may not be tailored to specific datasets. Incorporating domain knowledge remains challenging, as diagnostic features vary across anatomical regions and pathologies.

While many unsupervised methods require contrast-enhanced images for comparison, our approach operates without such reference images, addressing a critical gap in medical image enhancement. This capability is particularly significant for developing robust, generalizable methods that work across different imaging modalities and clinical settings. Our approach is valuable in emergency and resource-constrained environments where high-quality reference data may be unavailable. Furthermore, by reducing the potential for bias introduced by reference image selection, our method promotes consistent and reliable enhancement across diverse patient populations and imaging conditions.

\section{The Proposed Method}
\label{proposedmethod}

\subsection{Proposed Preprocessing}

\subsubsection{Intensity Inversion}
\noindent Our pre-processing phase consists of three sequential steps. First, we generate an inverted normalized image $\tilde{X}$, defined as:
\begin{eqnarray} \tilde{X}_{\nu} = \overline{X} - X_{\nu} \label{eq3} \end{eqnarray}
where $X$ is the original image, $\overline{X}$ represents its mean gray value, and $\nu$ denotes the spatial $(x,y)$ location of a pixel.
By inverting the intensity distribution relative to the mean, we enhance features adaptively, depending on their intensity.

\subsubsection{Weighted Least Squares (WLS) Smoothing}
\noindent Next, we generate a smoothed version of the original image, denoted as $g_{min}$, by applying the nonlinear WLS filtering~\cite{Fattal2008}. This operation balances blurring and sharpening to enhance the quality of X-ray images. It effectively smooths relatively homogeneous regions while preserving and enhancing gradients that represent key structural features (e.g., edges and textures). Given an input image $X$, the goal is to derive a new image $g$ that satisfies this fine balance. Mathematically, optimization can be formulated as finding the $g$ function that minimizes the following expression:

\begin{eqnarray}
    g_{min} &=& \underset{g}{\operatorname{argmin}} \sum_{\nu} \Bigg( \left( g_{\nu} - X_{\nu} \right)^{2} 
    + \lambda \Bigg( w_{x,\nu} \left( \frac{\partial g}{\partial x} \right)_{\nu}^{2}
    + w_{y,\nu} \left( \frac{\partial g}{\partial y} \right)_{\nu}^{2}
    \Bigg) \Bigg)
\end{eqnarray}

\noindent The term $(g_{\nu} - X_{\nu})^{2}$ minimizes the difference between $X$ and $g$, while ${\Big(w_{x,\nu} {\Big(\frac{\partial g}{\partial x}\Big)}^{2}_{\nu} + w_{y,\nu}{\Big(\frac{\partial g}{\partial y} \Big)}^{2}_{\nu}\Big)}$
promotes smoothness by minimizing the partial derivatives of $g$. The parameter $\lambda$ controls the balance between these terms, i.e., increasing $\lambda$ produces progressively smoother images $g$.

The weights $w_{x,\nu}$ and $w_{y,\nu}$ determine the level of smoothness along each image dimension at pixel $\nu$ and are defined as in previous work \cite{Lischinski2006}:

\begin{eqnarray}
    w_{x,\nu} ={ \Big(\Big|\frac{\partial l}{\partial x}}\Big|_{\nu}^{\alpha} +  \varepsilon\Big)^{-1}\qquad
    w_{y,\nu} ={\Big(\Big|\frac{\partial l}{\partial y}\Big|_{\nu}^{\alpha} +  \varepsilon \Big)^{-1},}
\end{eqnarray}

\noindent where $l$ is the log-luminance channel of the input image $X$ \cite{McCann2008}. The parameter $\alpha$ (in the range $1.2 < \alpha < 2.0$) determines the sensitivity of the low-frequency components (LFCs) to the log-luminance of the input image $X$. We set $\alpha$ to 1.2 to ensure that details and edges are well-preserved after smoothing. The parameter $\varepsilon$ (set to 0.0001) is a small constant that ensures numerical stability.

\subsubsection{Contrast Enhanced Image Generation}
\noindent An enhanced image $E$ is then obtained by combining $g_{min}$ with the inverted image $\tilde{X}$:

\begin{eqnarray} E = \tilde{X}+g_{min}. \label{eq4} \end{eqnarray}

\noindent By integrating the inverted image $\tilde{X}$ with $g_{min}$, we effectively enhance image contrast in high-frequency regions, while simultaneously preserving smoothness in homogeneous areas of the original image $X$. Image borders often display distinct pixel intensity variations across adjacent regions. To accentuate these differences and broaden the dynamic range, we introduce the inverted image $\tilde{X}$ specifically in these border regions. The pixel values from the inverted image contribute to an expanded intensity range, allowing for improved emphasis on subtle features and fine details within these critical areas.

\subsubsection{Final Pre-processed Image}
\noindent In the final pre-processing step, $E$ is used as a pixel-wise divisor to the original image:

\begin{eqnarray} X_{p_{\nu}} = \frac{X_{\nu}}{E_{\nu}}. \label{eq5} \end{eqnarray}

\noindent This pixel-wise division creates an inverse relationship between $E$ and the preprocessed image $X_{p}$. High-intensity regions in $E$ attenuate corresponding areas in $X_{p}$, reducing the dominance of naturally bright features. Conversely, low-intensity regions in $E$ amplify corresponding areas in $X_{p}$, enhancing subtle edges. This adaptive intensity modulation preserves strong structural features while amplifying weak transitions, resulting in balanced contrast enhancement across the image.

\begin{figure}[ht!]
    \centering
    \begin{minipage}[b]{0.22\textwidth}
        \centering
        \includegraphics[width=\linewidth]{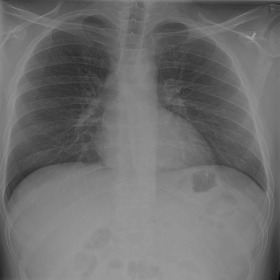}
        \caption*{\large{(a)}}
    \end{minipage}
    \hspace{0.02\textwidth} 
    \begin{minipage}[b]{0.22\textwidth}
        \centering
        \includegraphics[width=\linewidth]{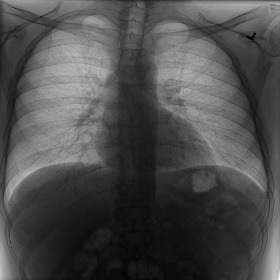}
        \caption*{\large{(b)}}
    \end{minipage}
    \hspace{0.02\textwidth} 
    \begin{minipage}[b]{0.22\textwidth}
        \centering
        \includegraphics[width=\linewidth]{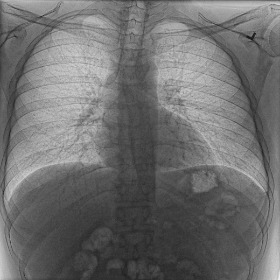}
        \caption*{\large{(c)}}
    \end{minipage}\\
    
    \hspace{0.02\textwidth} 
    \begin{minipage}[b]{0.22\textwidth}
        \centering
        \includegraphics[width=\linewidth]{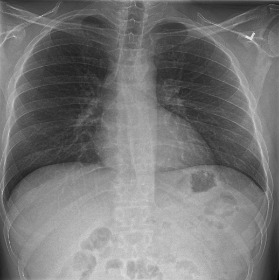}
        \caption*{\large{(d)}}
    \end{minipage}
    \hspace{0.02\textwidth} 
    \begin{minipage}[b]{0.22\textwidth}
        \centering
        \includegraphics[width=\linewidth]{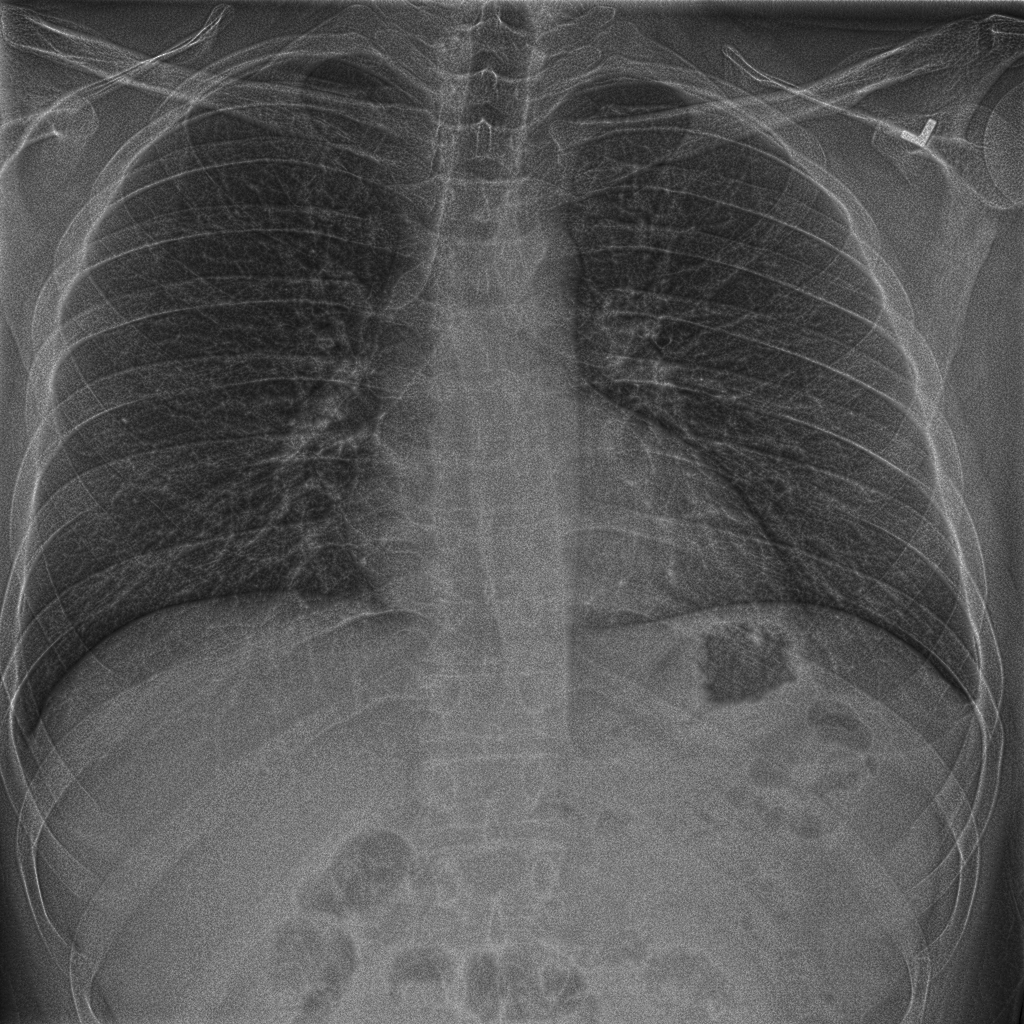}
        \caption*{\large{(e)}}
    \end{minipage}
    \caption{Illustration of pre-processing stages: (a) Input image $X$, (b) inverted image $\tilde{X}$ (Eq. 1), (c) combined image $E$ (Eq. 4),  (d) final pre-processing image $X_p$ (Eq. 5), and (e) HFC image $Y$ (Eq. 6).}
    \label{fig1}
\end{figure}

Also, we obtain a high frequency components (HFC) image $Y$ by subtracting the LFCs from the preprocessed X-ray image $X_p$, i.e.,

\begin{equation}
Y = X_{p} - {g_p}_{min}.
\label{eq6}
\end{equation}
where ${g_p}_{min}$ is obtained by WLS-filtering of $X_p$. Fig.~\ref{fig1} illustrates the above described pre-processing stages. \\

\subsection{Proposed Learning Architecture}
\noindent The proposed model is built upon a \textit{U-Net architecture}, which is well suited to handle limited training data \cite{Ding2019}, \cite{Hatamizadeh2019} and features an additional embedded internal GL. The network inputs the original image $X$ and the HFC image $Y$ and attempts to generate an enhanced output after each iteration (see Fig.~\ref{fig4}).

\subsubsection{U-Net Backbone with Guidance Layer}
Each layer in the U-Net decoder is connected to its corresponding encoder layer, allowing for the integration of high-resolution information from the encoder. However, X-ray images often lack sufficient high-frequency details for effective training. To address this, we embed supplementary GLs within the decoder. These layers provide additional information that helps enhance subtle details that are otherwise difficult to detect.

The GLs are trained simultaneously with the rest of the network. As the network learns, these layers adjust to the data characteristics, directly impacting the activations of other layers. These adaptive GLs manage the flow of information within the network. After each training step, they supply more refined information (according to Eq.~\ref{eq6}), enhancing the network’s ability to emphasize vertebral contrast in X-rays.\newline

\subsection{Proposed Guidance Layer}
Our embedded GL is designed as an integral part of the main network and is trained jointly with the other layers. This implies that the GL can directly influence the activation patterns of the other layers, allowing it to modulate the information flow and representation within the network. As a result, the GL can help preserve information by providing additional constraints such as spatial consistency, structural regularization, and contrast enhancement and regularization that encourage the model to learn more structured and informative representations. 

\subsubsection{Structure of Guidance Layer}
An internal GL is generally more effective than an external one because it is not affected by the ``noisy'' statistics of training examples from external datasets. This allows us to harness overfitting beneficially, similar to internal learning processes. Our proposed GL layer consists of three components: 1) A \textit{Leaky rectified linear unit} (Leaky ReLU), 2) a \textit{convolution layer}, and 3) a \textit{max pooling layer}. The Leaky ReLU activation function is used to retain information from negative pixel values, the convolution layer extracts detailed features, and the max pooling layer adjusts the spatial dimensions to align with the network's architecture (Fig.~\ref{fig4}). The feature maps produced by the GL maintain the spatial location and shape of the spinal region, even in the deepest layers with lower-resolution images. As previously noted, after each training step, the GL is updated with refined information (as described in Eq.~\ref{eq6}) derived from the enhanced input, thereby improving the contrast of the vertebrae.

\begin{figure*}[ht!]
\centering
\begin{minipage}[b]{1.0\textwidth}

\includegraphics[width=\linewidth]{Fig2.jpg} 
 \end{minipage}
 \caption{Diagram of proposed model. Cropped preprocessed image and HFC image (obtained by Eq.~\ref{eq6}) are fed into the network and GL, respectively; extracted features from GL are concatenated with the network's convolutional feature maps in the decoder module. After each training step, the network and GL are updated due to enhanced output and HFC which are fed into the next iteration.}
\label{fig4}
\end{figure*}

\subsection{Iterative Training Procedure}
\noindent Our model employs an iterative training process, where the entire training procedure is repeated a number of times. Each subsequent iteration leverages the refined output from the previous one. We experimented with three training iterations and found that optimal results were achieved after only two iterations. In the first iteration, the model significantly enhances overall image visualization and reveals hidden structures within the spinal region. The second iteration builds upon these results, identifying and capturing more complex and intricate structures, thereby improving accuracy. A third iteration generally produces a noisier output. Therefore, the training process is formulated as:

\begin{equation}
         {X^{t+1}_{p}} =\phi(X^{t}_{p}, Y^{t}),      t = 0, 1
         \label{eq7}
   \end{equation} 

where $\phi$ denotes the U-Net operation, $X^{t}_{p}$ and $Y^{t}$ are the cropped outputs from the $t$-th training step, and $X^{0}_{p} = X_{p}$, $Y^{0} = Y$ are the pre-processed image and the derived HFC image, respectively. These are used as network inputs for the next training step, i.e., for $t = 1$.

\subsection{Enhancement-Oriented Loss Function}
We propose a loss function designed to capture enhanced image details. The overall loss is divided into two sublosses.

\subsubsection{\textbf{Structural Boundary Similarity}}
We first introduce a gradient-correlation loss between the processed and HFC images to improve accuracy and emphasize the borders of local structures. \textit{Gradient correlation} (GC) \cite{Penney1998} has been widely used in medical image registration and has demonstrated its effectiveness in improving boundary accuracy in medical images~\cite{Hiasa2018}. Let $A$, $B$ denote two images whose average values are $\overline{A}$ and $\overline{B}$, respectively. The GC measure of the two images is obtained by taking the average of the \textit{normalized cross-correlation} (NCC) values across both horizontal and vertical directions of the images, i.e.,

\begin{eqnarray}
       GC(A,B) = \frac{1}{2}[ {NCC(\nabla_{x}{A}, \nabla_{x}{B})}
       + {NCC(\nabla_{y}{A},\nabla_{y}{B})}],
       \label{eq:eq8}
       \end{eqnarray}
where the NCC of the images is given by
\begin{equation}
       NCC(A,B)=\frac{ \sum_{i}\sum_{j} {(A_{ij} - \overline{A})}^2 {(B_{ij}- \overline{B})}^2}
       {\sqrt{ \sum_{i}\sum_{j} {(A_{ij} - \overline{A})}^{2} }  \sqrt{ \sum_{i}\sum_{j} {(B_{ij} - \overline{B})}^{2} }} 
        \end{equation} 
        
We define the gradient correlation loss between the two images of interest as

\begin{equation}
\centering
       \mathcal{L}_{1}^{t}(X^{t}_{p},Y^{t}) = {1-GC(X^{t}_{p},Y^{t}), {t=1,2}} 
       \label{eq10}
        \end{equation}
where $\mathcal{L}_{1}^{t}$ is the loss over the $t$-th training step.  
We obtain ${Y^{t}}$ by applying Eq. (\ref{eq6}) to the network output after each training step, i.e., for ${t=1,2}$. 

Minimizing this loss during training encourages the neural network to produce images that closely match the desired ones in terms of gradient correlation.

\subsubsection{\textbf{Feature Enhancement}}
Due to the lack of supervision from ground-truth data, we propose regularizing the training using information extracted from the input. From the network's last layer $\phi$, we collect all 128 feature maps and extract the most informative ones. These 128 feature maps consist of 64 network feature maps and 64 GL feature maps. Low-level feature maps capture image color, edges, contrast, and textures. In our method, we pre-train the network using a gradient correlation loss to help it learn spinal structures and contours that are not clearly visible in the images. This enables the network to effectively capture subtle contrast and detailed spatial information, thereby improving perceptual connectivity on low-level features. Combining low-level details with high-level semantics, as achieved in the last layer, yields superior results compared to relying solely on raw features from early layers.

We use informative \textit{entropy} as a criterion to identify and filter out redundant feature maps. (Recall that feature maps with low entropy are considered less informative.) However, entropy alone does not distinguish between meaningful information and noise. While entropy has its limitations, it offers a computationally efficient method for pruning redundant channels and retaining potentially useful ones. Combined with our learning strategy, which allows the network to progressively learn complex features, entropy serves as a foundation for feature-based loss. This enables the network to focus on extracting intricate details within these established structures.

To calculate the information entropy of an image, we use the following formulation:

\begin{equation}
         {\mathcal H(Z)} = - \sum^{255}_{i=0}{p(z_{i}){\rm {log}}({p(z_{i})})}
         \label{eq11}
   \end{equation}
   
where $Z$ represents the discrete random variable of an image grayscale value, and ${p(z_{i})}$ is the probability that $Z = z_{i}$. To calculate the information entropy for a feature map, we first map the pixel values of each feature map ${c_{l}}$ to the range $0\mbox{--}1$ and group them into $K$ bins to reduce computational cost. We use ${x_{a,b}}$ to denote a pixel in the feature map.

The entropy of the $l$-th feature map is then calculated according to Eq.~(\ref{eq11}) as follows:

\begin{eqnarray}
    {\mathcal{H}(c_{l})} &=& -\sum_{i=0}^{K-1} \Bigg( \frac{\sum_{a=1, b=1}^{n', m'} f(c(x_{a,b})=i)}{n' \times m'} \Bigg) \times \log \Bigg( \frac{\sum_{a=1, b=1}^{n', m'} f(c(x_{a,b})=i)}{n' \times m'} \Bigg)
    \label{eq12}
\end{eqnarray}
where $f$ is the indicator function, and $n'$ and $m'$ are the height and width, respectively, of the feature map.

To exploit these properties for structural enhancement, we first select the feature map with the highest entropy, designating it as the most informative channel, ${{c}^{*}}$. The loss function guides the network to enhance regions with a more diverse range of pixel intensities by encouraging the model to focus on features with higher entropy. This results in improved visibility of subtle details, enhanced contrast in important structures, and an overall more informative image representation.

By following these steps, we can effectively leverage the low-level features of the neural network to enhance the visibility of desired structures in the spinal region.

The feature-based loss is defined as the MSE loss between the most informative feature map and the network output, namely:

\begin{equation}
      \mathcal{L}^{t}_{2} = {\| X^{t}_{p} - {c}^{*}}  \|_2^2
      \label{eq8}
      \end{equation} 

Thus, the overall objective function is obtained as the sum of the individual loss functions, i.e., 
\begin{equation}
 \centering
       \mathcal{L}= {\mathcal{L}_{1}} + {\mathcal{L}_{2}}
         \end{equation}

The training strategy for the model is implemented gradually over each training step. Initially, during the first 200 epochs, the model is trained according to $\mathcal{L}_{1}$ to allow the network to learn the basic data features. This is because $\mathcal{L}_{1}$ helps the network learn the structural boundaries of the HFC image. In the subsequent 200 epochs for larger datasets and 25 epochs for smaller datasets, after the network has acquired the data features, we apply the overall $\mathcal{L}$ loss. During this phase, $\mathcal{L}_{2}$ extracts the most detailed feature map from the network's last layer, leveraging the features already learned.\\

\section{Experiments and Analysis}
\label{experimentsanalysis}

\subsection{Datasets}
\label{experimentsanalysis1}
\noindent We assessed the performance of our proposed model on the following four datasets to demonstrate its capabilities and robustness across diverse data statistics:

\begin{enumerate}
\item
The JSRT dataset~\cite{Shiraishi2000} contains 247 grayscale X-ray images, each with a resolution of  $2048 \times 2048$ pixels and a 12-bit depth per pixel.

\item
The Montgomery County X-ray dataset~\cite{Jaeger2014} contains 138 posterior-anterior X-ray scans, each with a resolution of $4020 \times 4892$ or $4892 \times 4020$ pixels and a 12-bit depth per pixel.

\item
The  ChestX-ray14 dataset~\cite{wang2017} contains 112,120 grayscale X-ray images collected from 30,805 patients. Each image is of size $1024 \times 1024$ pixels and an 8-bit depth per pixel.

\item
The CheXpert dataset~\cite{Irvin2019} is a large open-source collection of multi-label chest X-ray images, with varying dimensions and an 8-bit depth per pixel. It contains 224,316 chest radiographs from 65,240 unique patients.
\end{enumerate}

\begin{figure}[ht!] 
    \centering
    \begin{minipage}[b]{0.22\textwidth} 
        \centering
        \includegraphics[width=\linewidth]{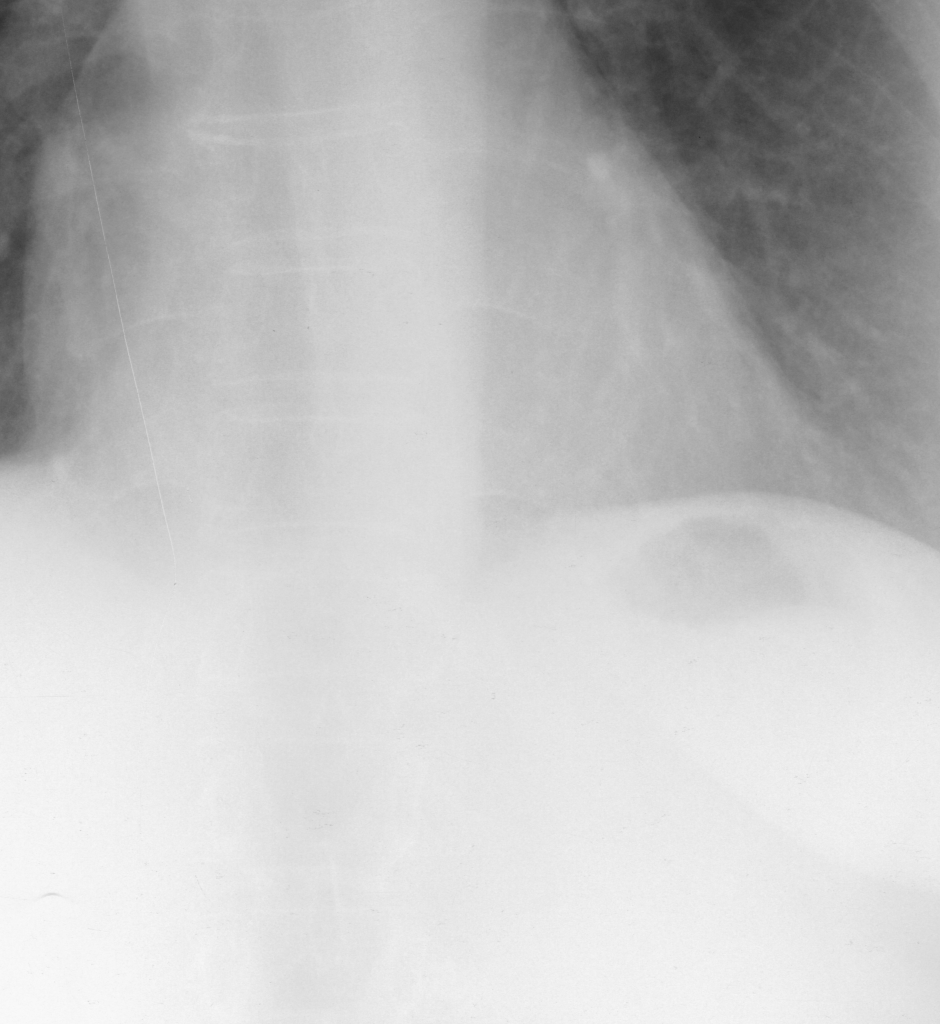} 
        \caption*{\large{(a)}}
    \end{minipage}
    \hspace{0.02\textwidth} 
    \begin{minipage}[b]{0.22\textwidth}
        \centering
        \includegraphics[width=\linewidth]{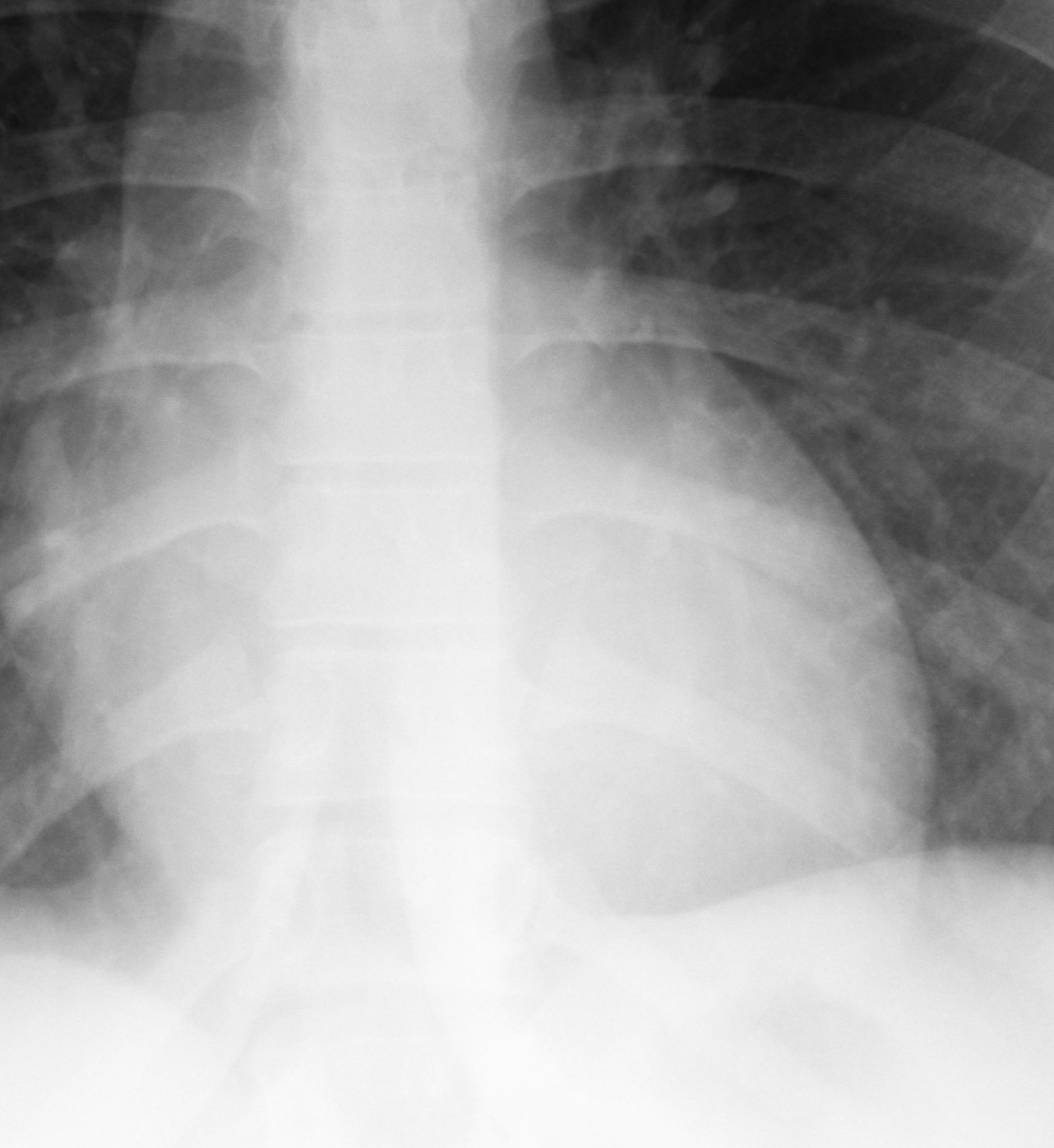}
        \caption*{\large{(b)}}
    \end{minipage}
    \hspace{0.02\textwidth}
    \begin{minipage}[b]{0.22\textwidth}
        \centering
        \includegraphics[width=\linewidth]{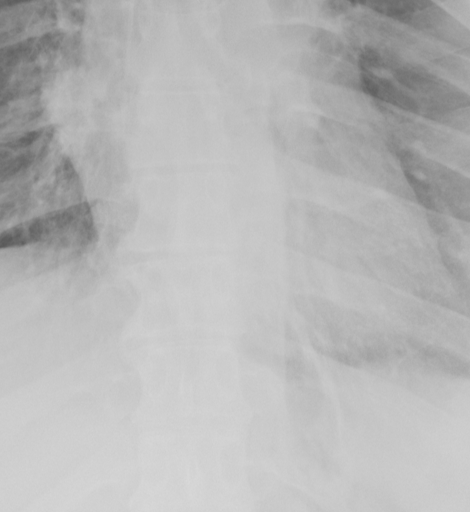}
        \caption*{\large{(c)}}
    \end{minipage}
    \hspace{0.02\textwidth}
    \begin{minipage}[b]{0.22\textwidth}
        \centering
        \includegraphics[width=\linewidth]{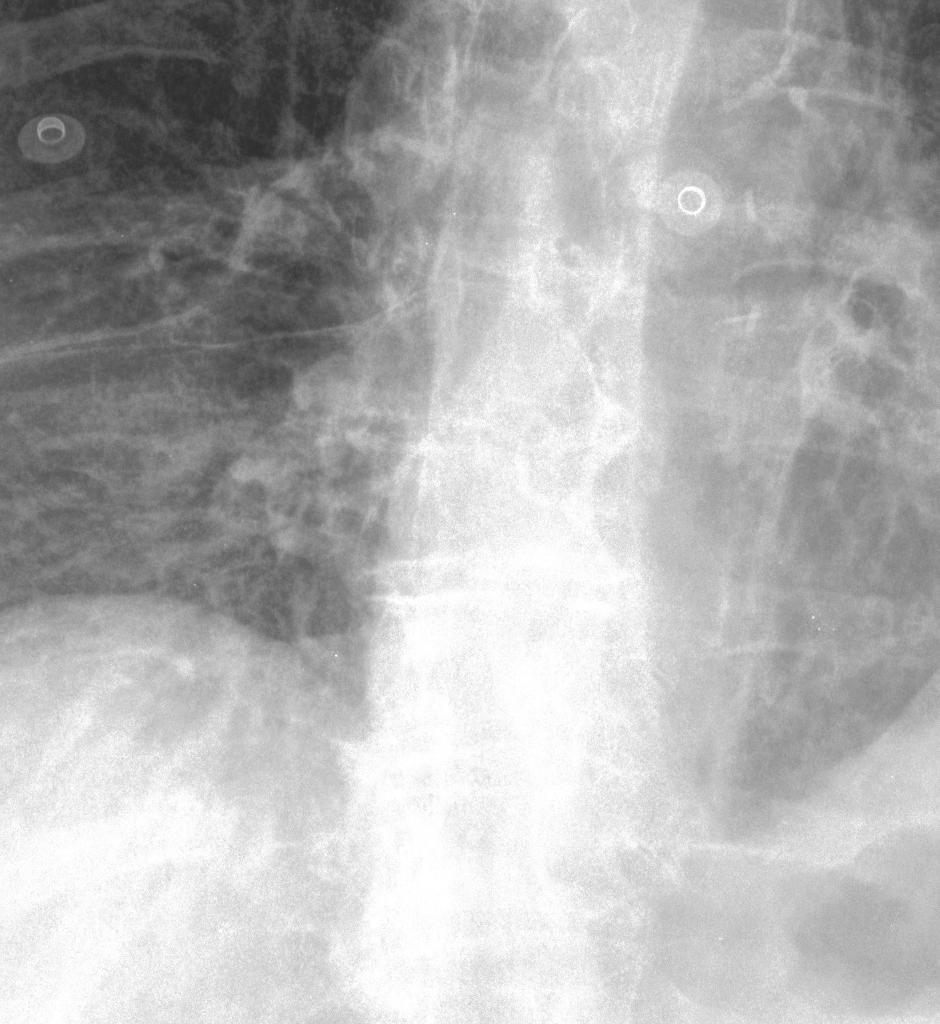}
        \caption*{\large{(d)}}
    \end{minipage}
    \caption{Image diversity across four datasets: (a) JSRT, (b) Montgomery County X-ray, (c) NIH ChestX-ray14, and (d) CheXpert}
    \label{fig3}
\end{figure}

\noindent Fig.~\ref{fig3} shows image samples from the above datasets.

\subsection{Training XVertNet}
We developed a patch-wise U-Net-based architecture that processes two $512 \times 512$-pixel regions of interest (ROIs) for each image; one centered on the upper spinal cord and another on the lower spinal cord. As shown in Fig.~\ref{fig3}, our datasets exhibit diverse spatial characteristics, with images varying in dimensions (e.g., $2048 \times 2048$, $4020 \times 4892$, and $1024 \times 1024$ pixels) and bit depths (8-bit and 12-bit). This variability in image features poses a significant challenge for creating a robust enhancement model.

\begin{figure*}[p]
\centering
\setlength{\tabcolsep}{2pt}
\renewcommand{\arraystretch}{1.2} 

\begingroup
\setkeys{Gin}{height=0.12\textheight,keepaspectratio}
\begin{tabular}{>{\centering\arraybackslash}m{0.05\textwidth} 
                >{\centering\arraybackslash}m{0.187\textwidth}
                >{\centering\arraybackslash}m{0.187\textwidth}
                >{\centering\arraybackslash}m{0.187\textwidth}
                >{\centering\arraybackslash}m{0.187\textwidth}
                >{\centering\arraybackslash}m{0.187\textwidth}}

& \text{JSRT} & \text{Montgomery} & \text{NIH} & \text{CheXpert}\\

\multirow{2}{*}[0.7pt]{\rotatebox{90}{\text{ Original}}} &
\includegraphics[width=\linewidth]{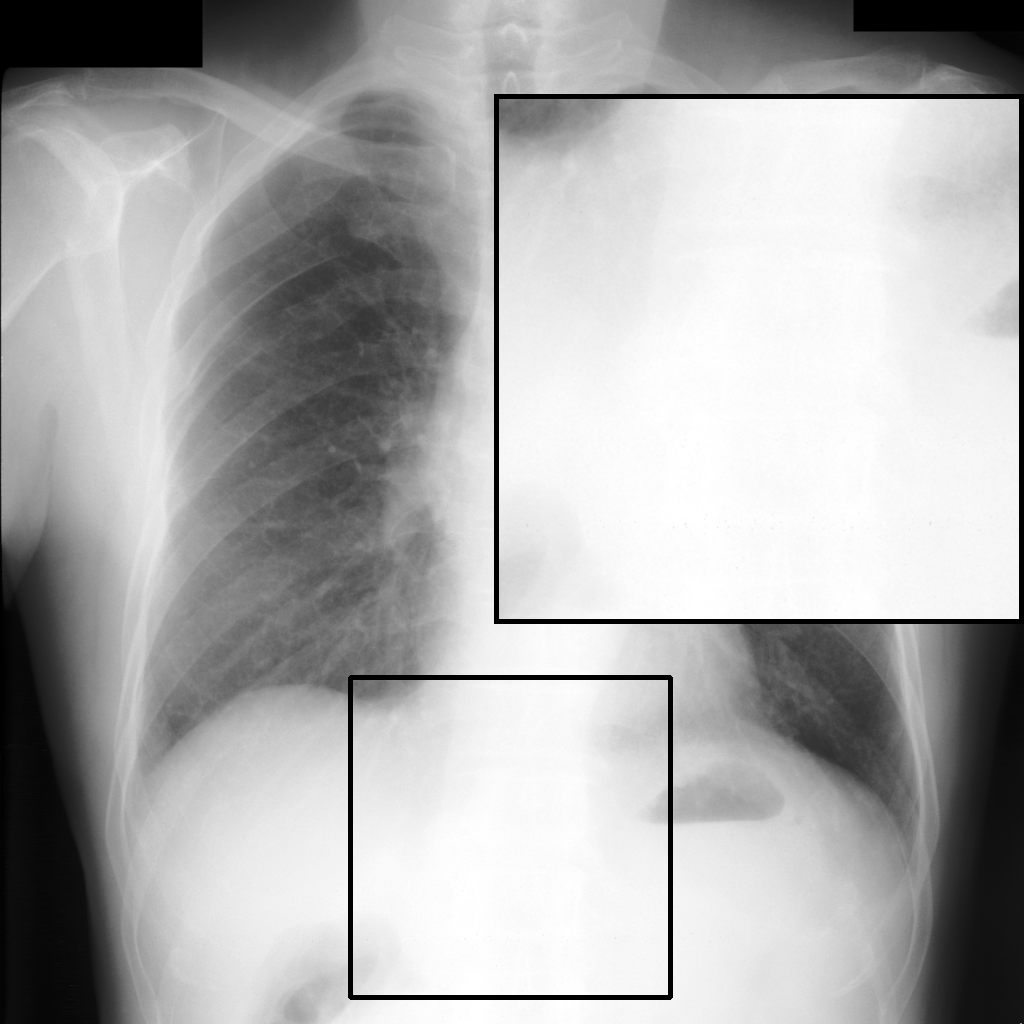} &
\includegraphics[width=\linewidth]{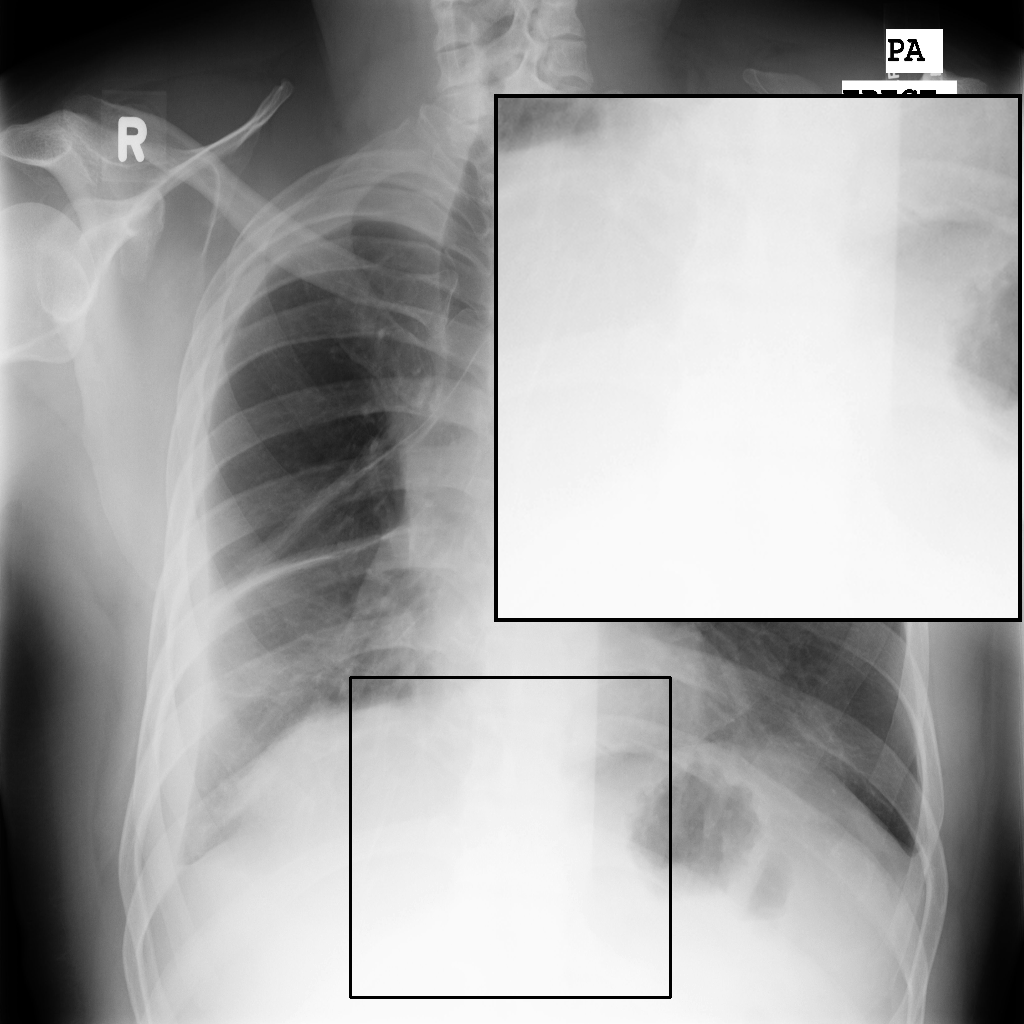} &
\includegraphics[width=\linewidth]{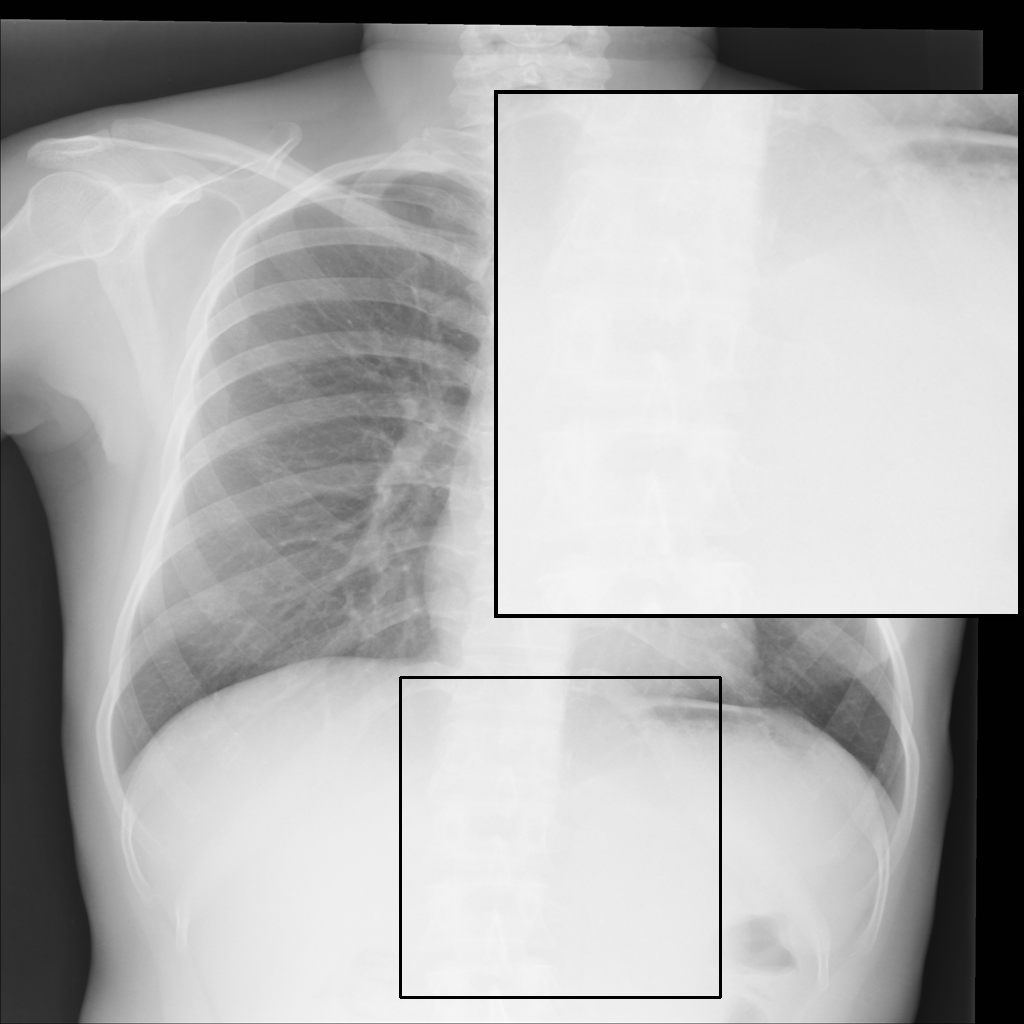} &
\includegraphics[width=\linewidth]{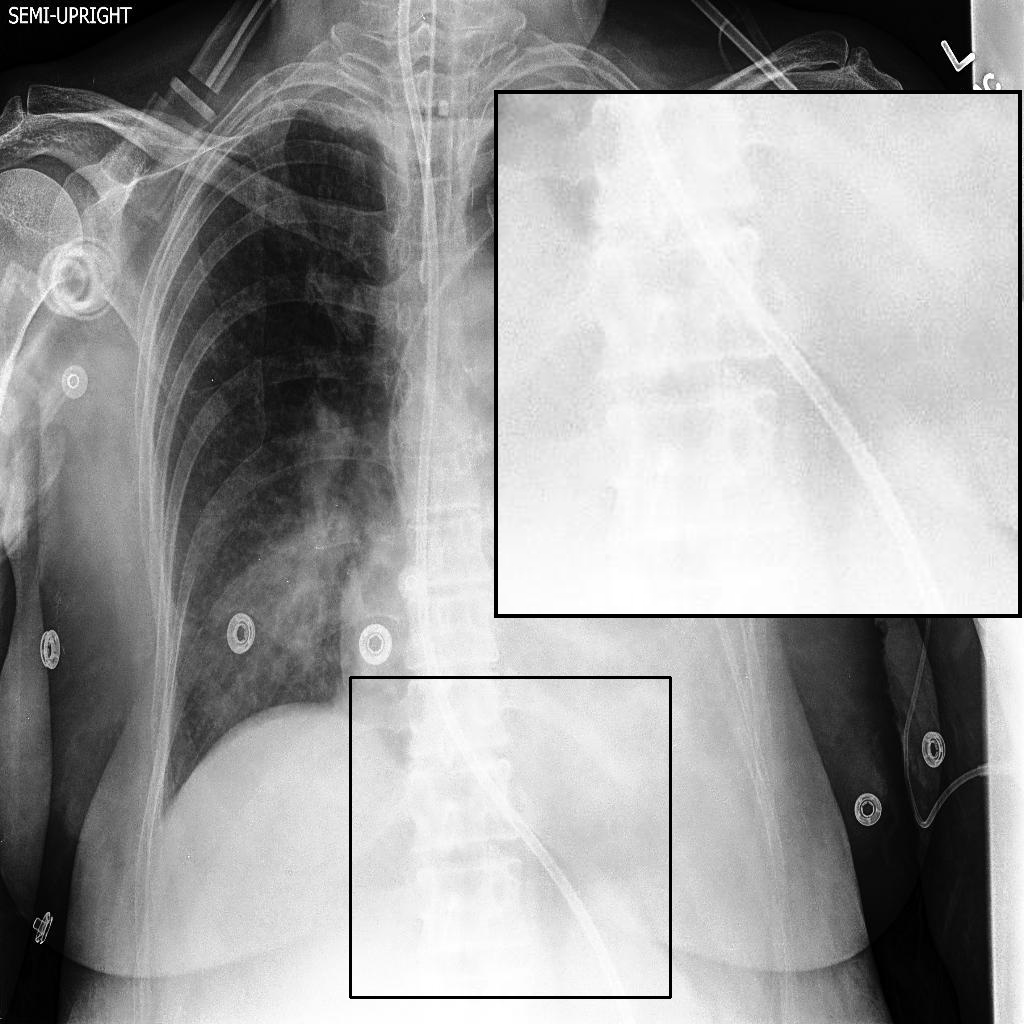}  \\

\multirow{2}{*}[1em]{\rotatebox{90}{\text{CLAHE \cite{Pisano1998}}}} &
\includegraphics[width=\linewidth]{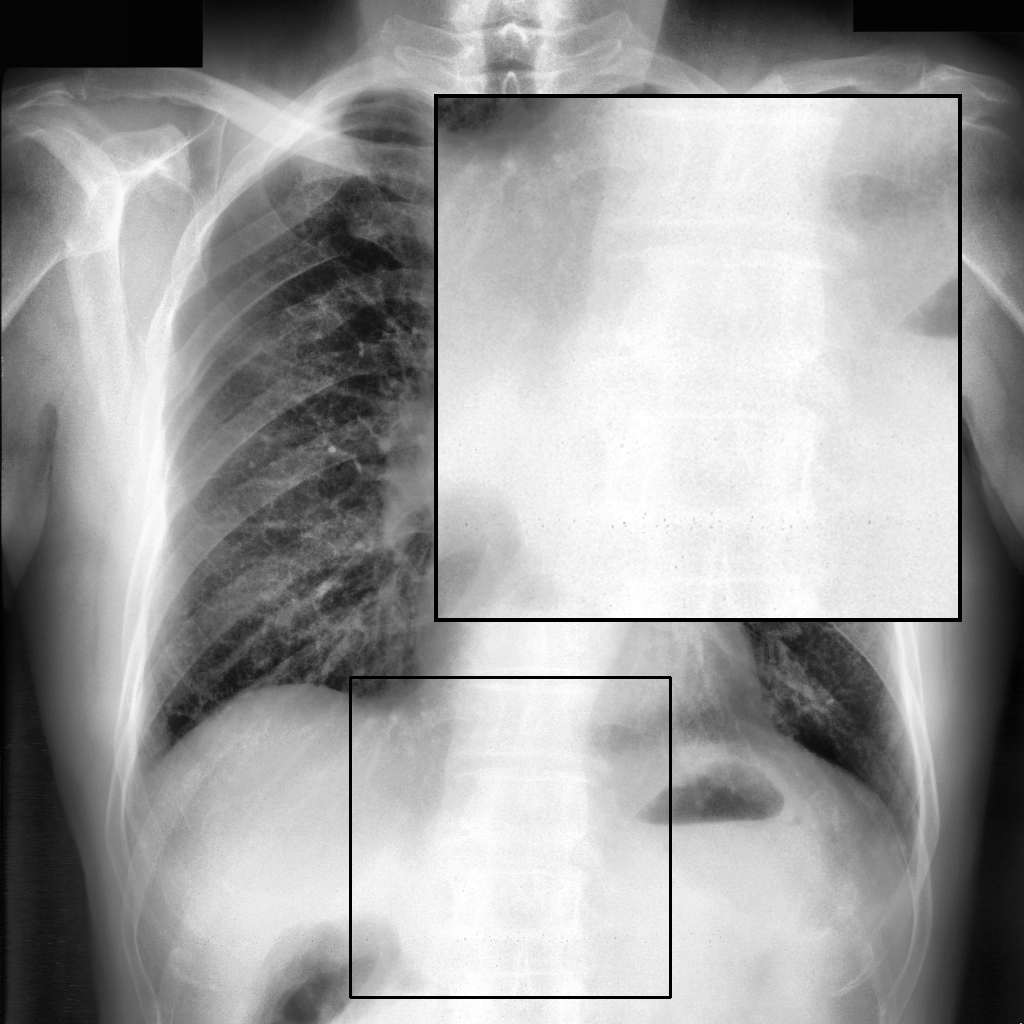} &
\includegraphics[width=\linewidth]{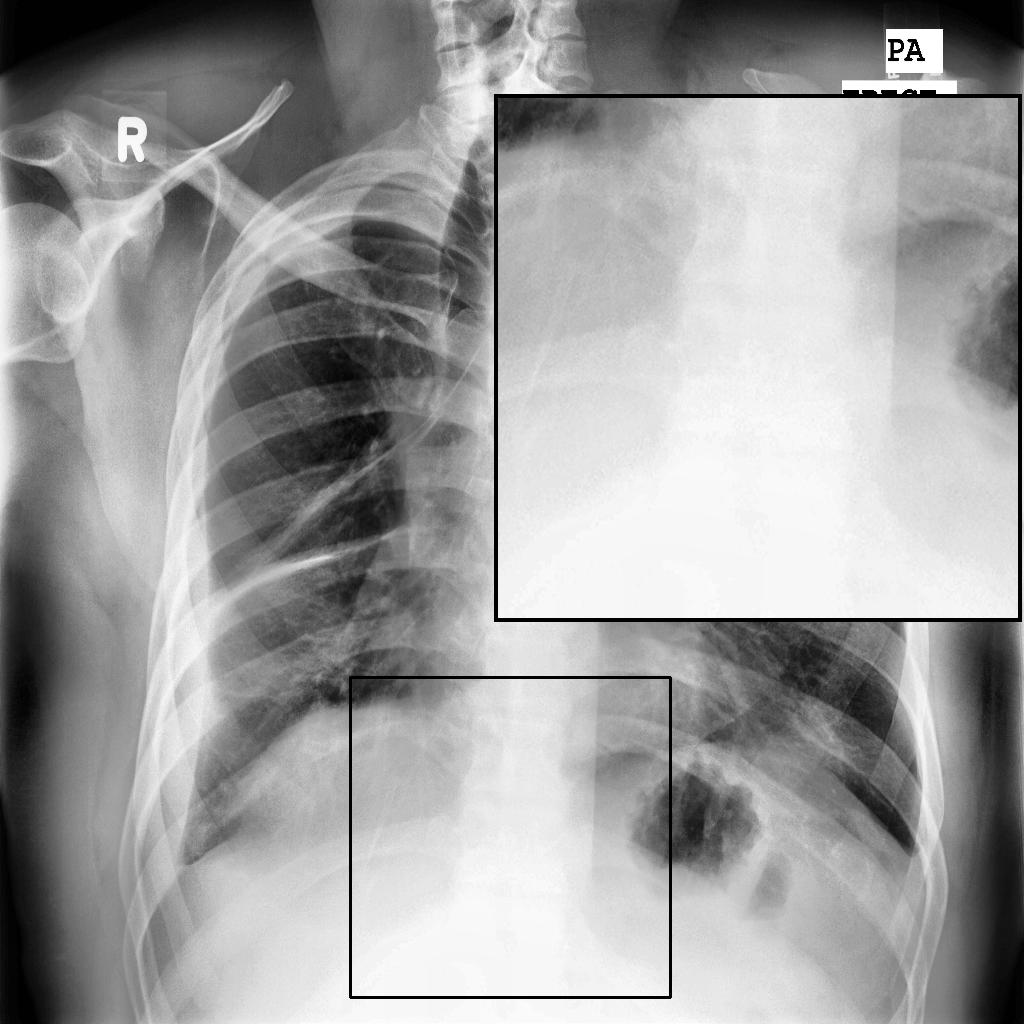} &
\includegraphics[width=\linewidth]{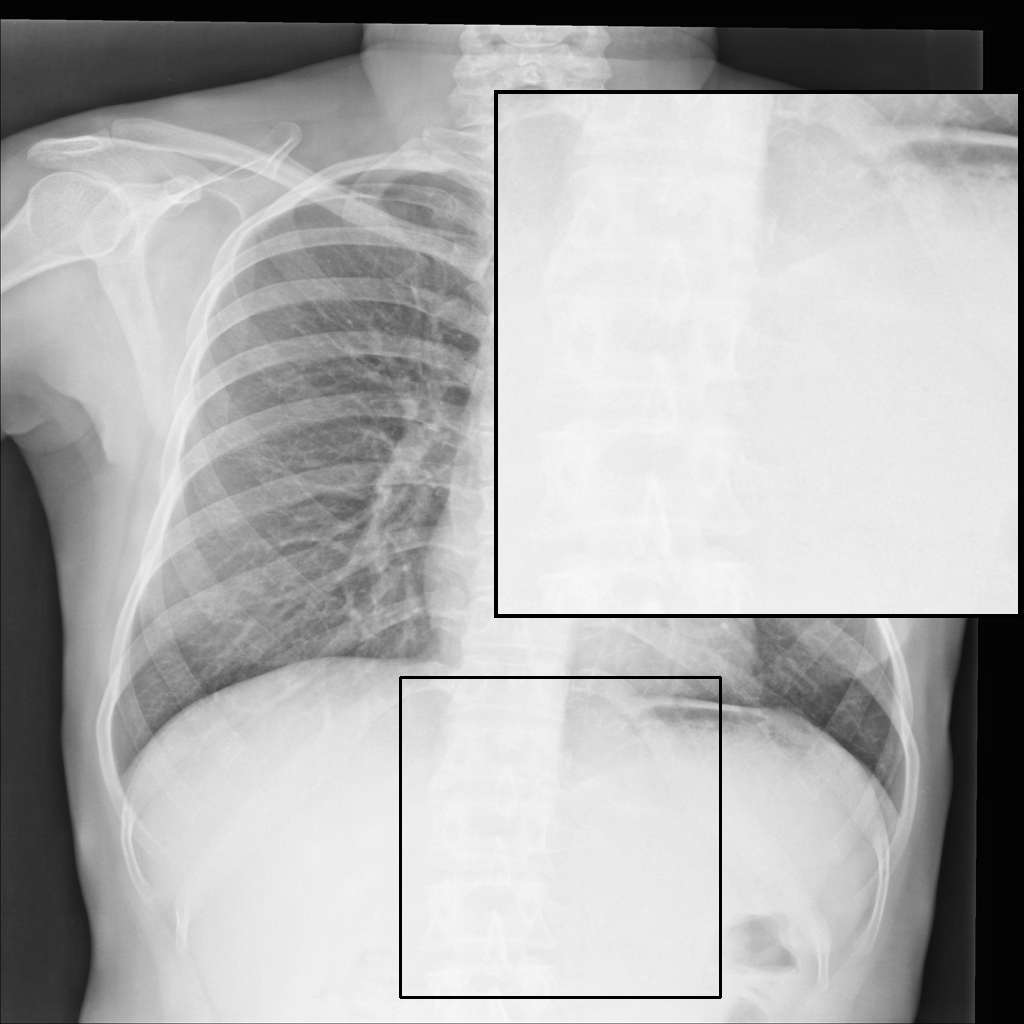} &
\includegraphics[width=\linewidth]{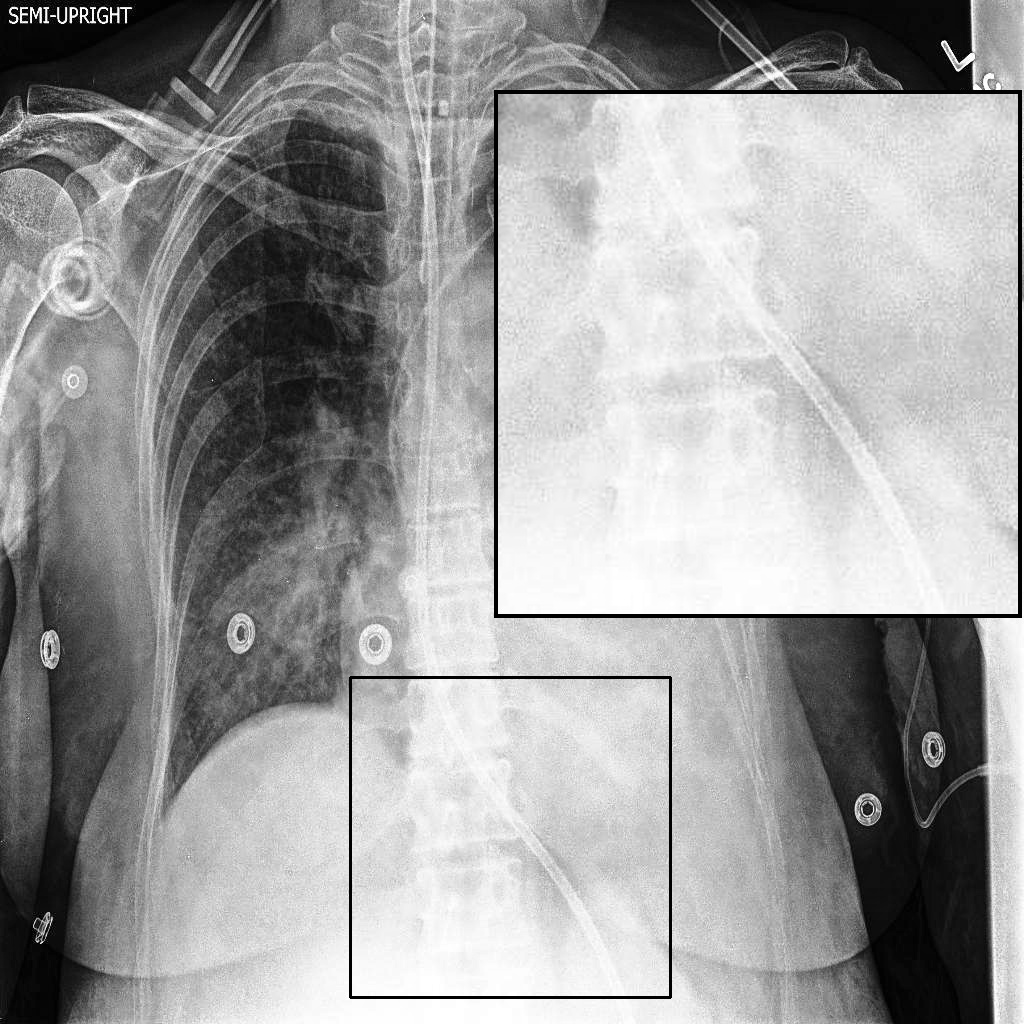} \\

\multirow{2}{*}[0.07\columnwidth]{\rotatebox{90}{\text{Farbman \textit{et al.} \cite{Fattal2008}}}} &
\includegraphics[width=\linewidth]{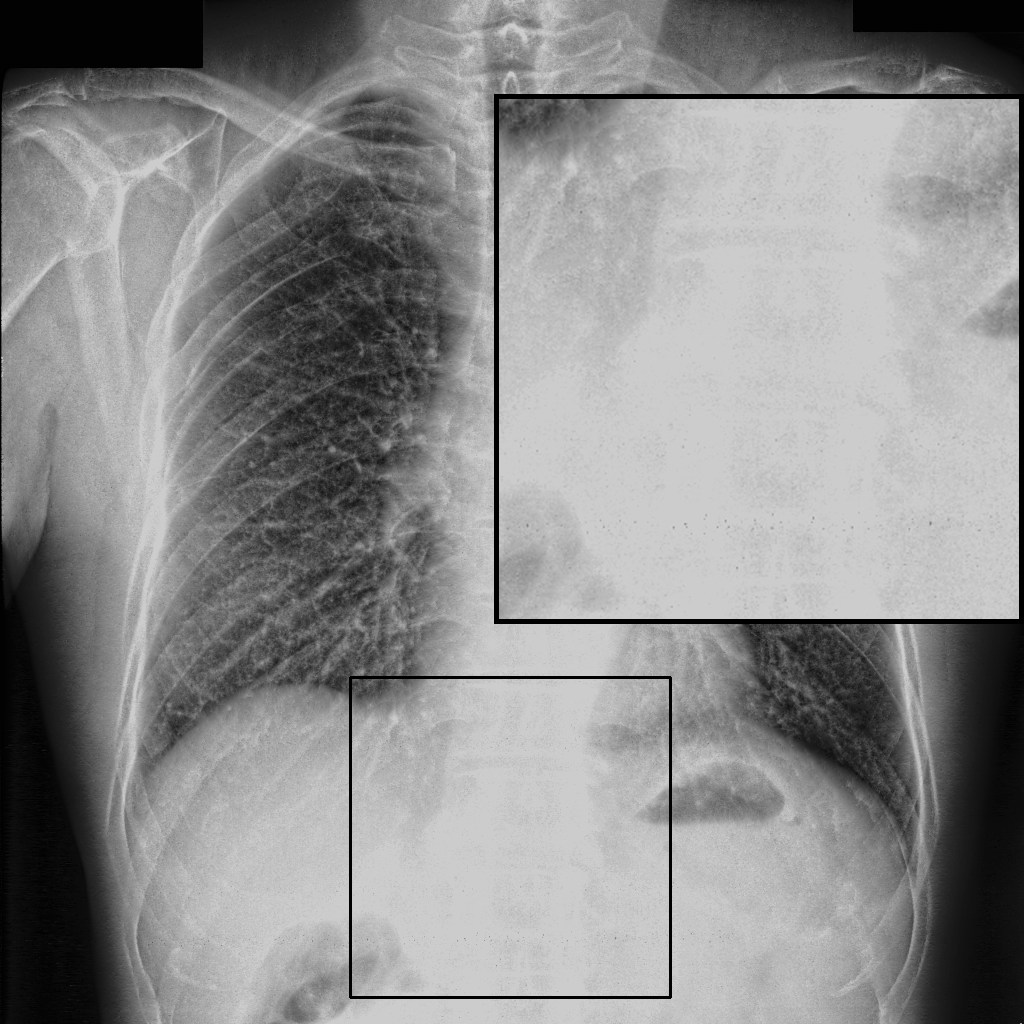} &
\includegraphics[width=\linewidth]{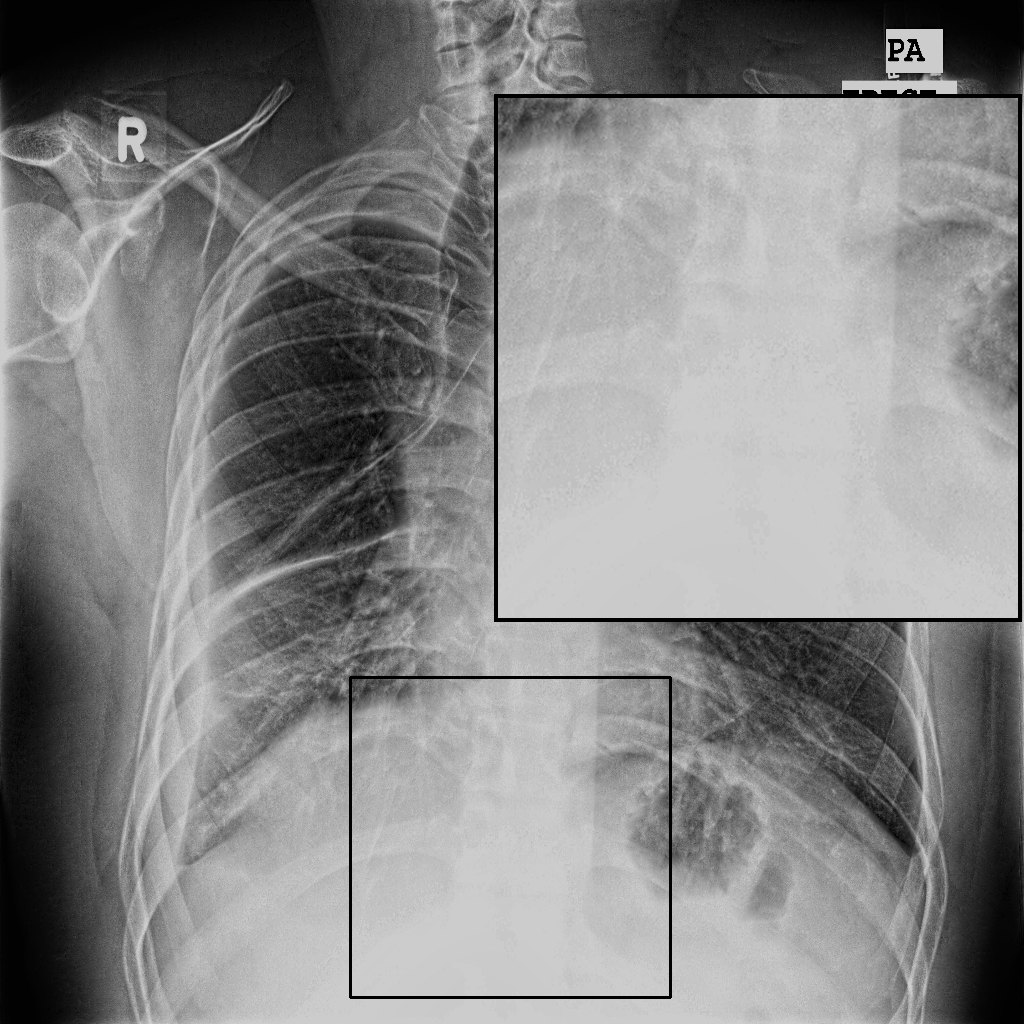} &
\includegraphics[width=\linewidth]{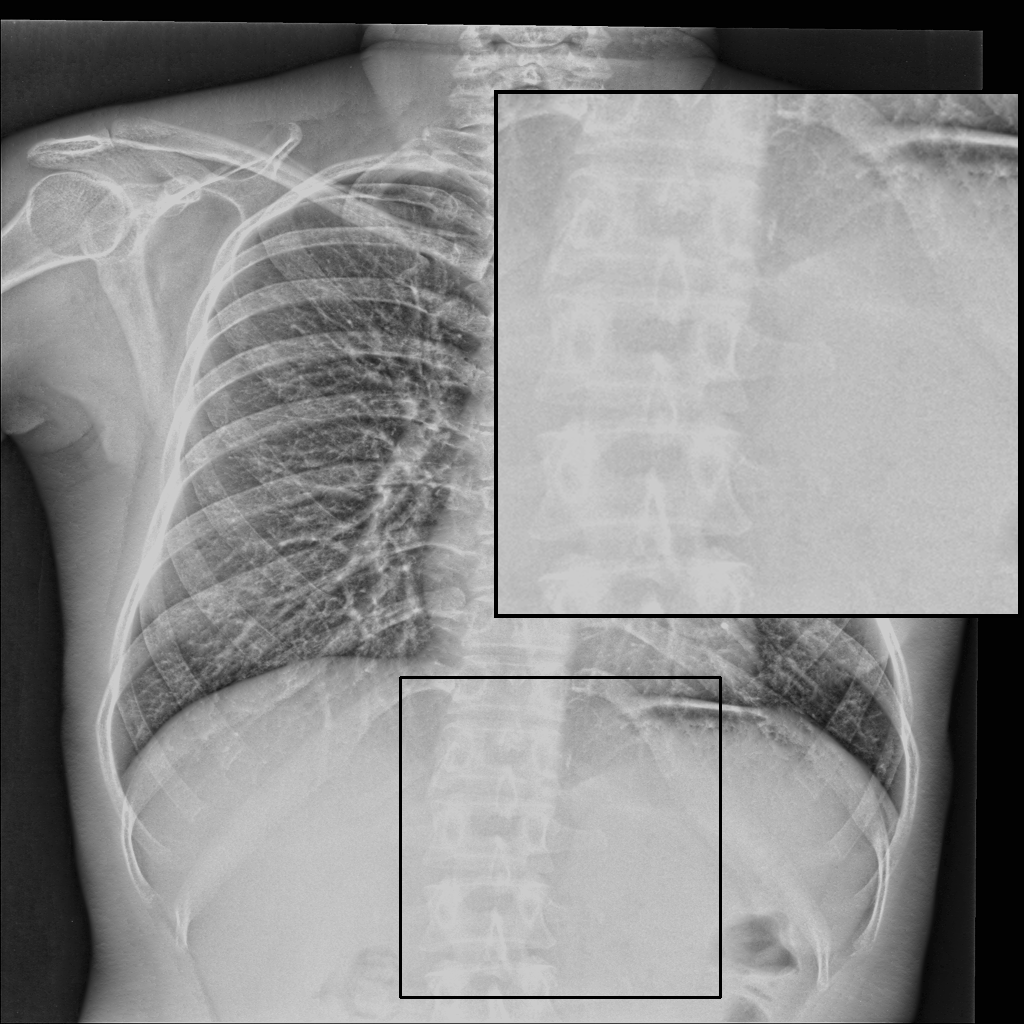} &
\includegraphics[width=\linewidth]{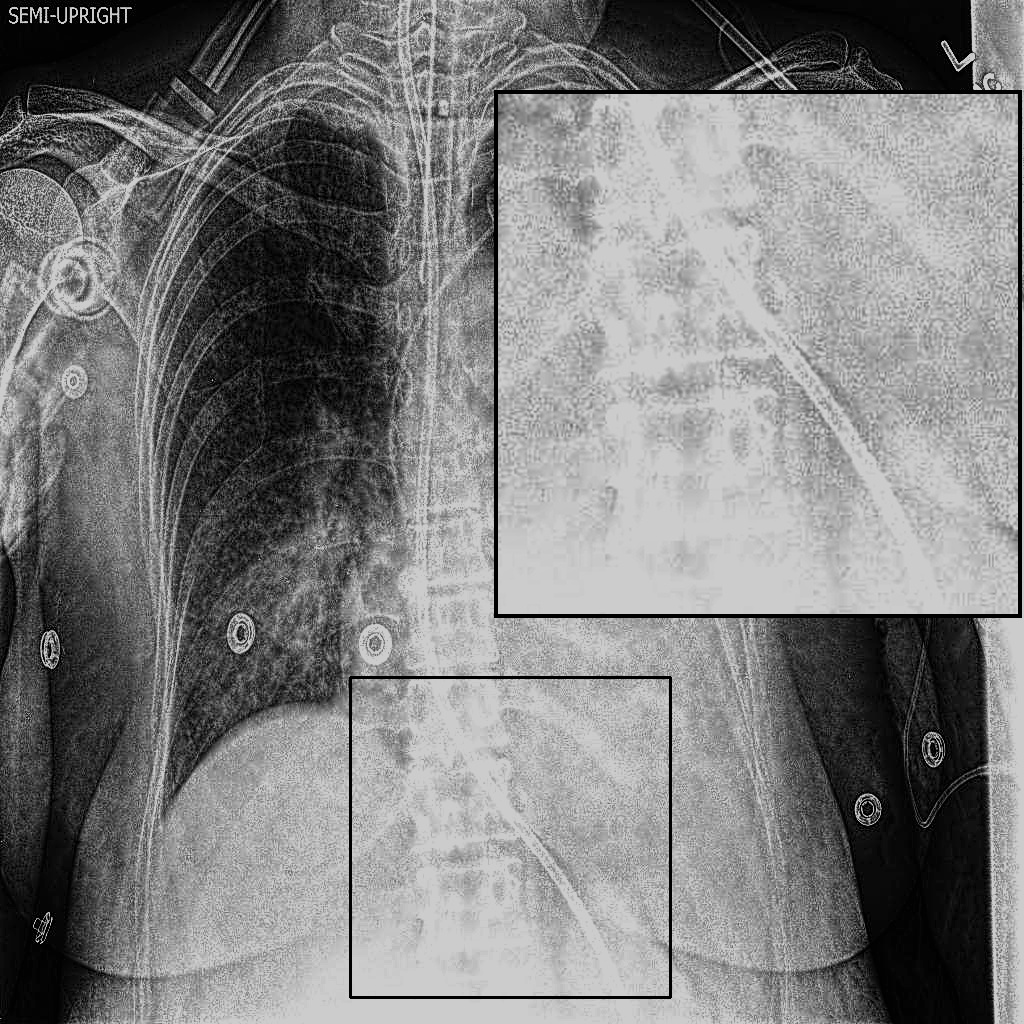}  \\

\multirow{2}{*}[1em]{\rotatebox{90}{\text{ZSSR \cite{Shocher2018}}}} &
\includegraphics[width=\linewidth]{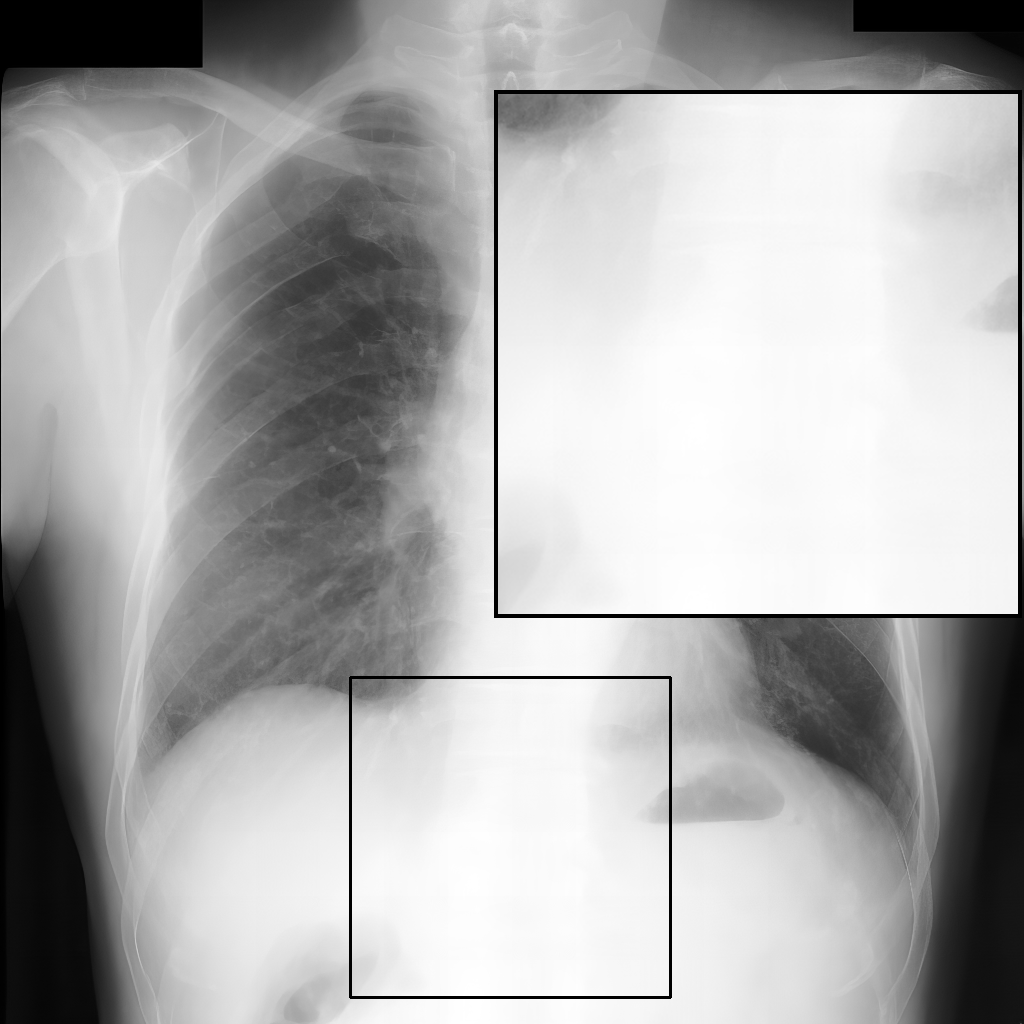} &
\includegraphics[width=\linewidth]{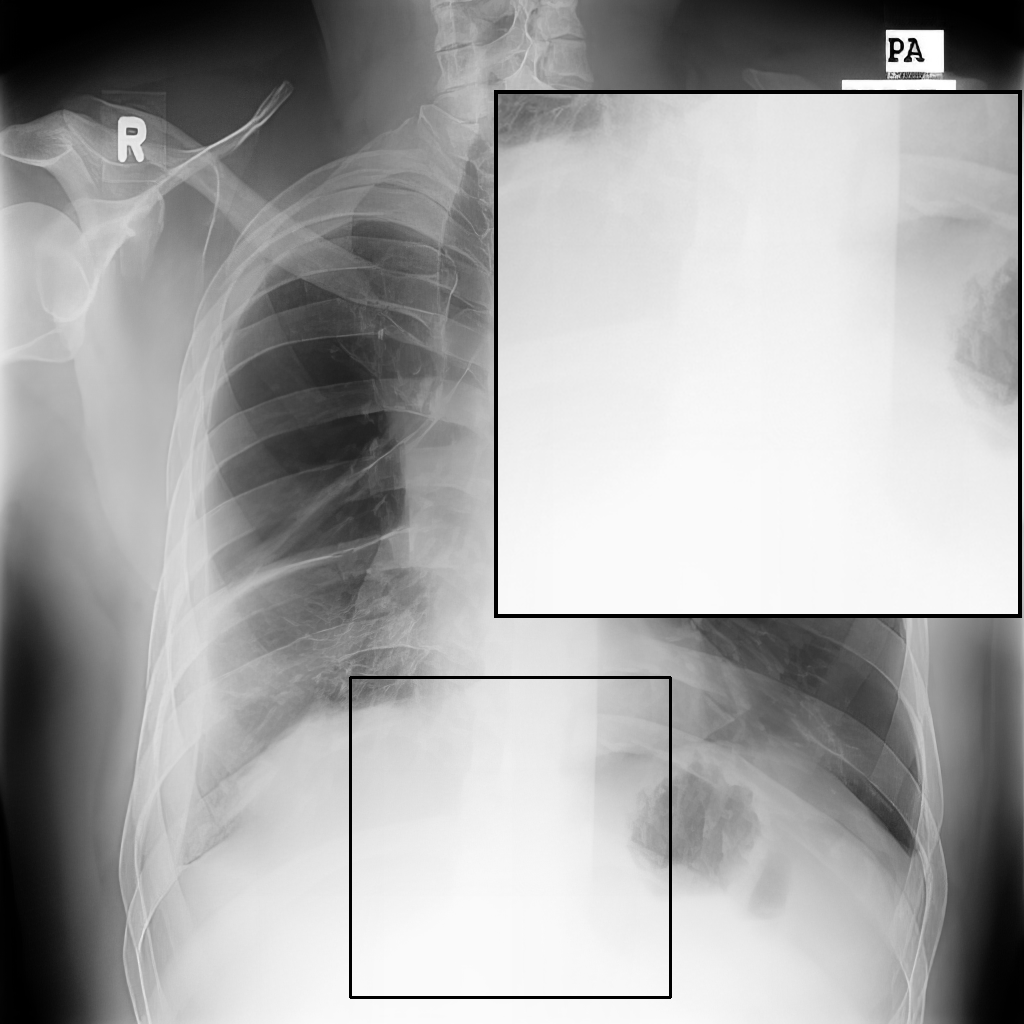} &
\includegraphics[width=\linewidth]{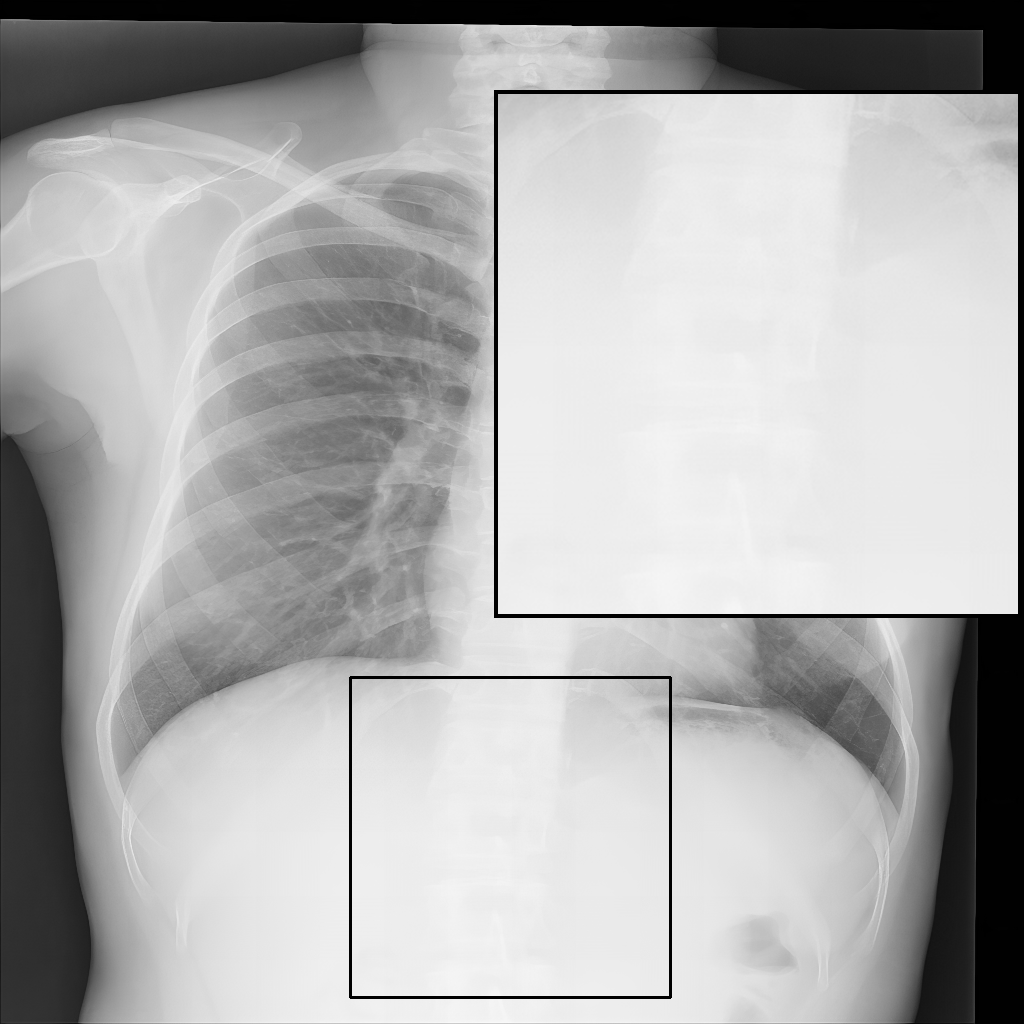} &
\includegraphics[width=\linewidth]{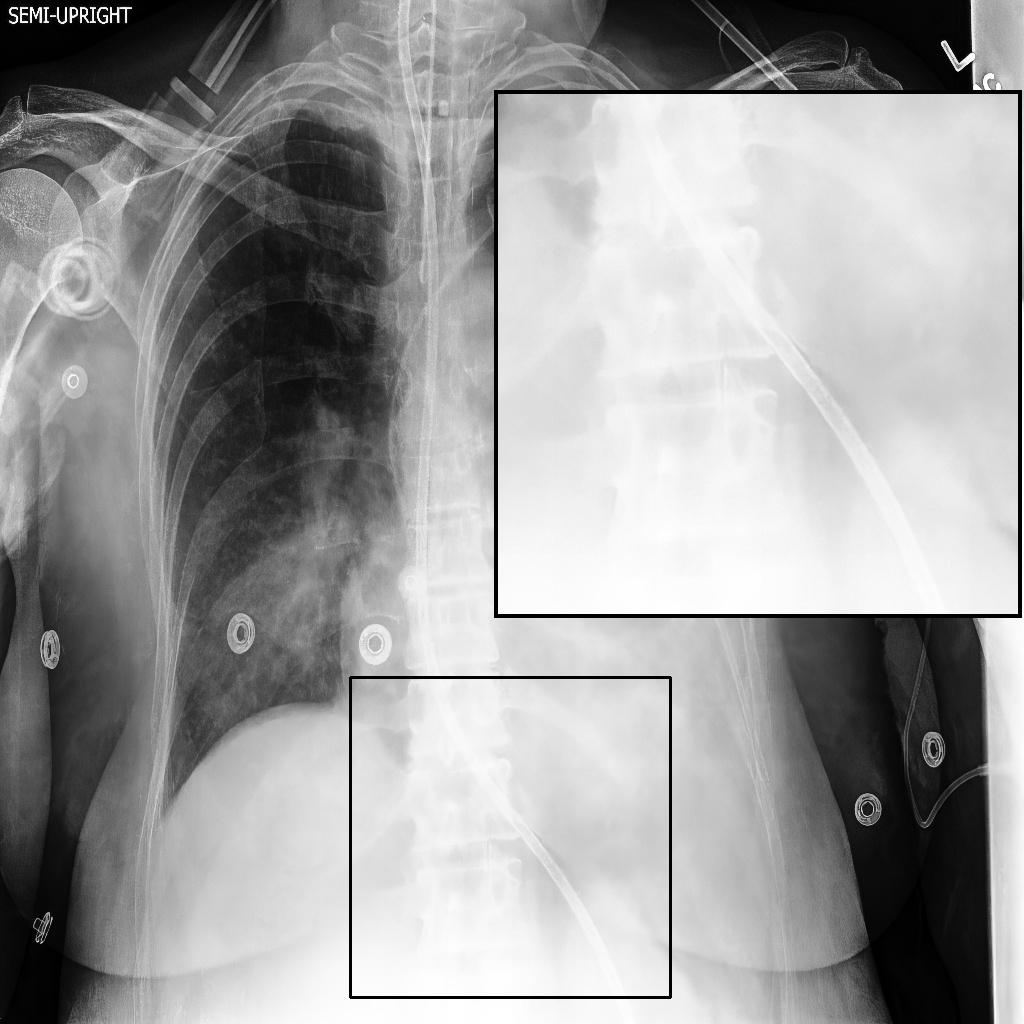}  \\

\multirow{2}{*}[0.07\columnwidth]{\rotatebox{90}{\text{ Zero-DCE ~\cite{guo2020}}}} &
\includegraphics[width=\linewidth]{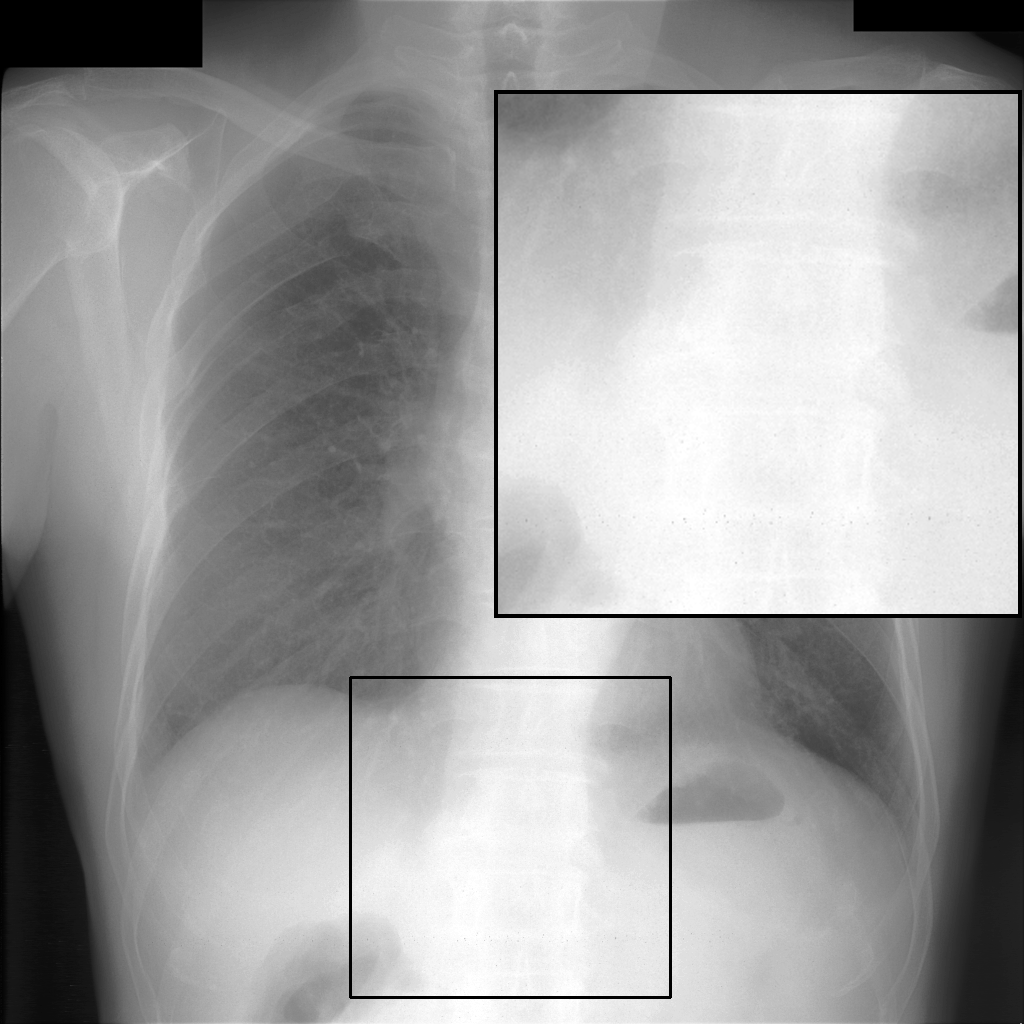} &
\includegraphics[width=\linewidth]{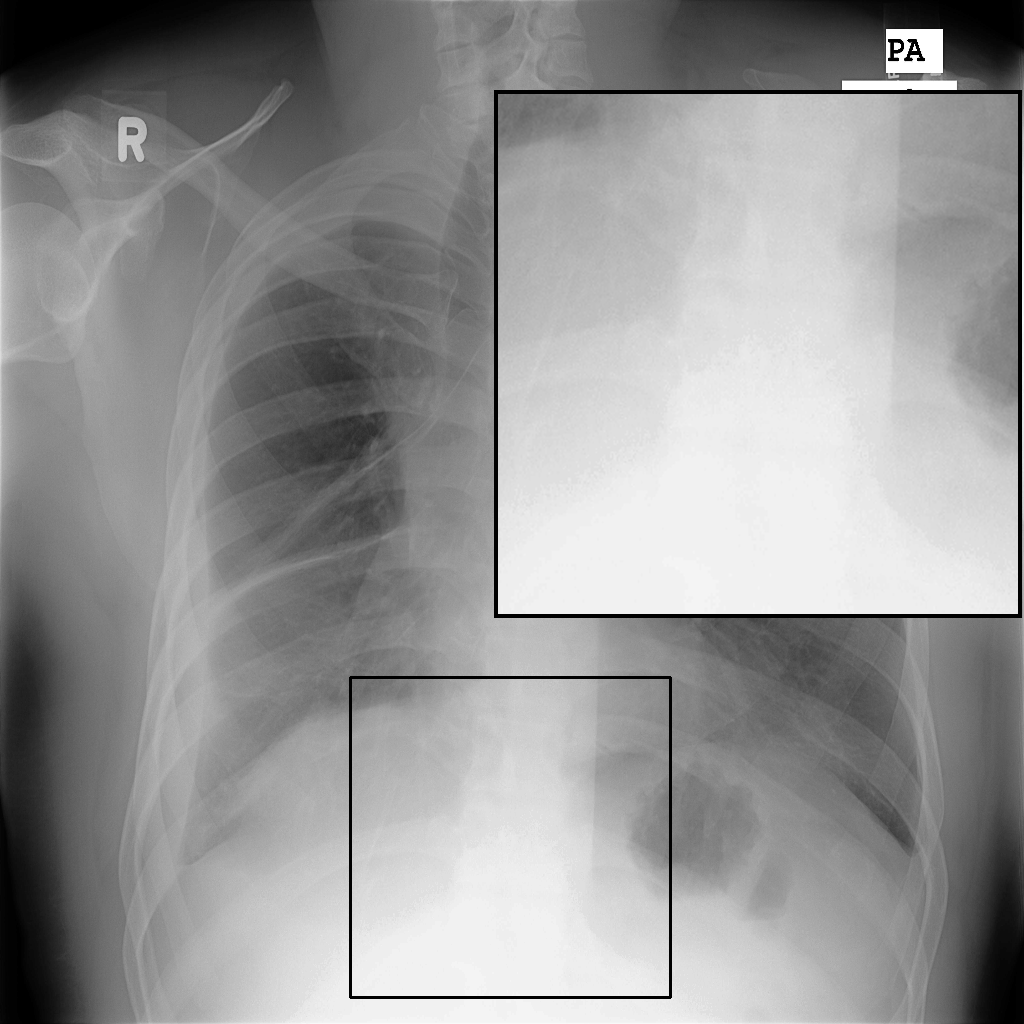} &
\includegraphics[width=\linewidth]{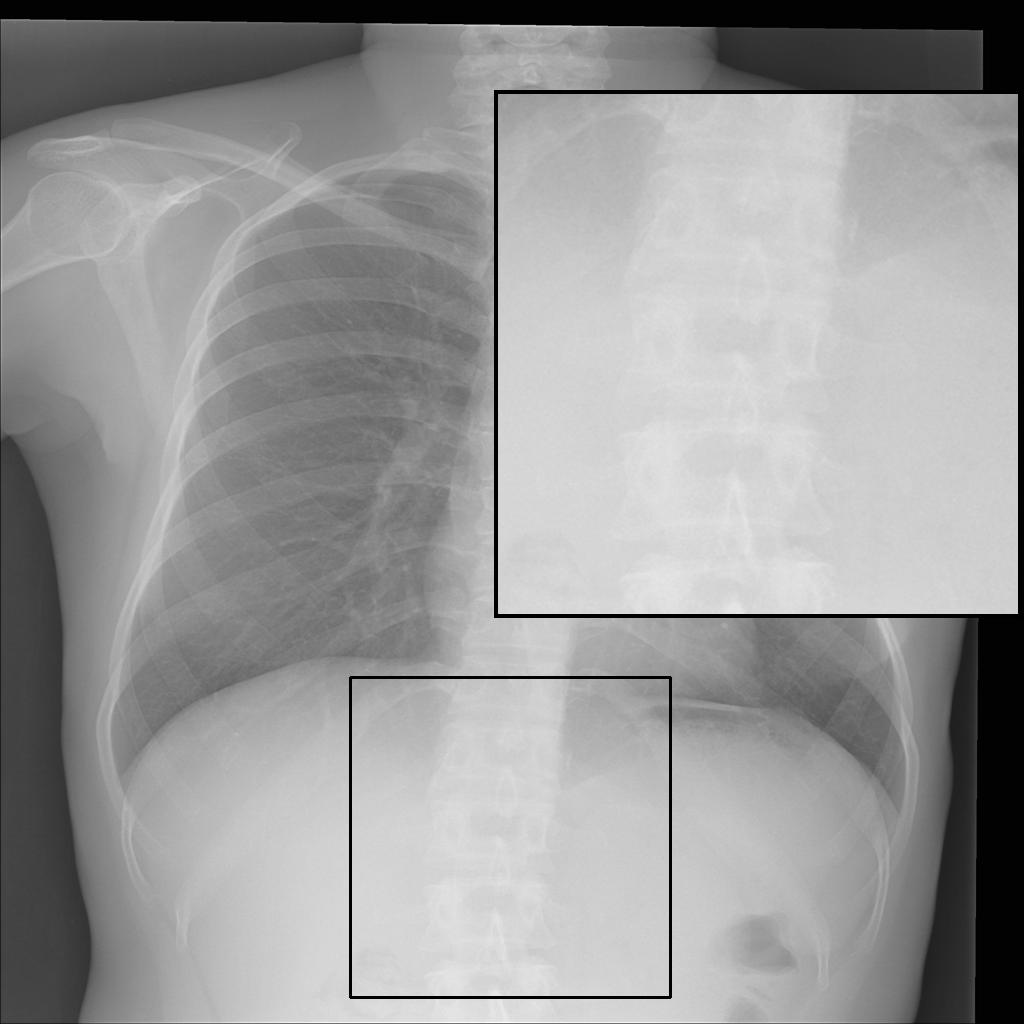} &
\includegraphics[width=\linewidth]{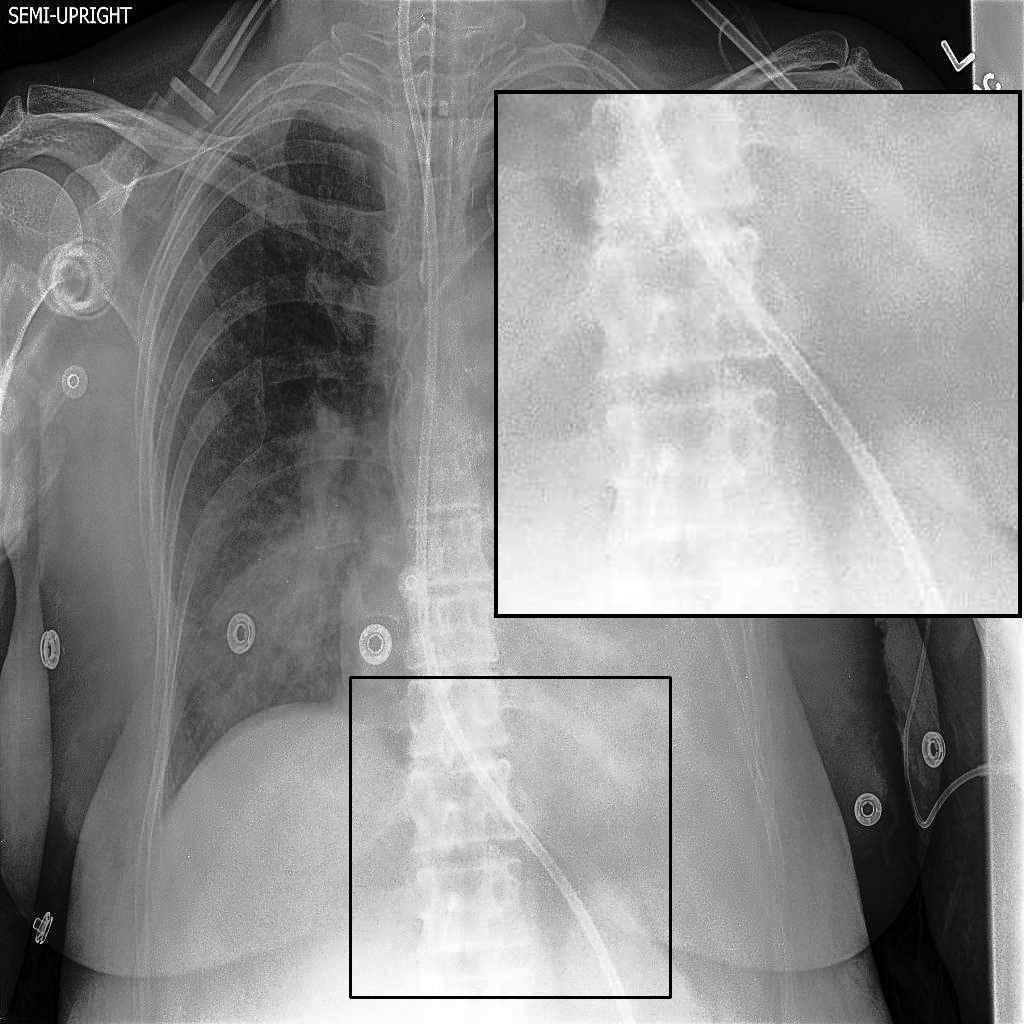}  \\

\multirow{2}{*}[0.07\columnwidth]{\rotatebox{90}{\text{ Madmad \textit{et al.} \cite{Madmad2021}}}} &
\includegraphics[width=\linewidth]{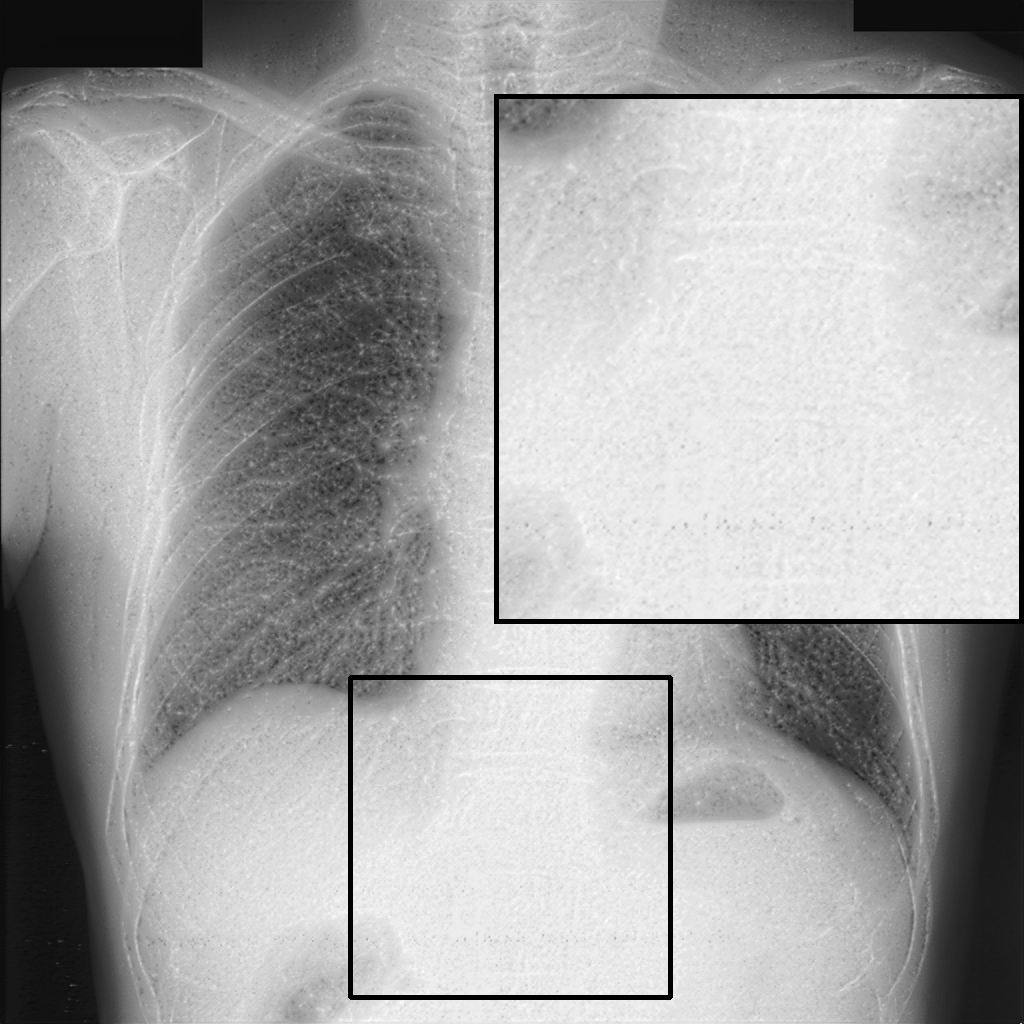} &
\includegraphics[width=\linewidth]{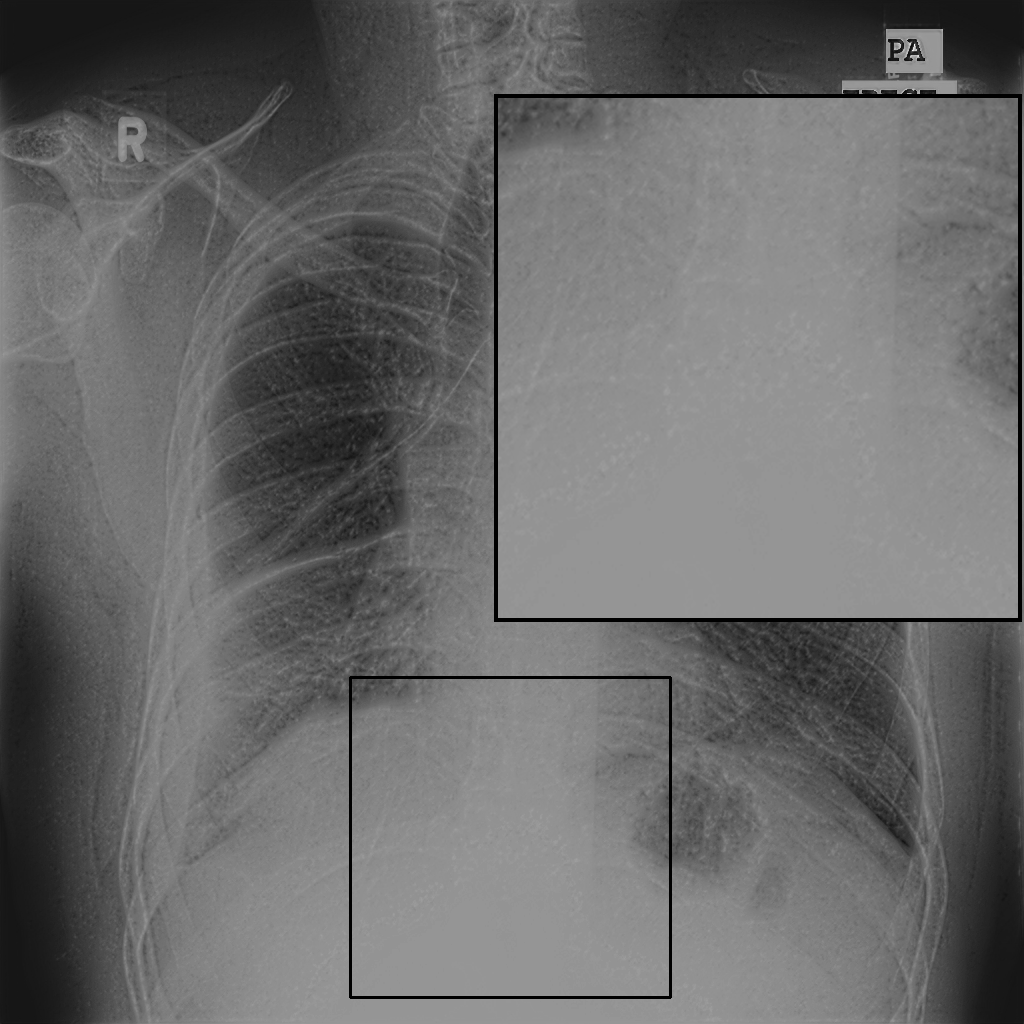} &
\includegraphics[width=\linewidth]{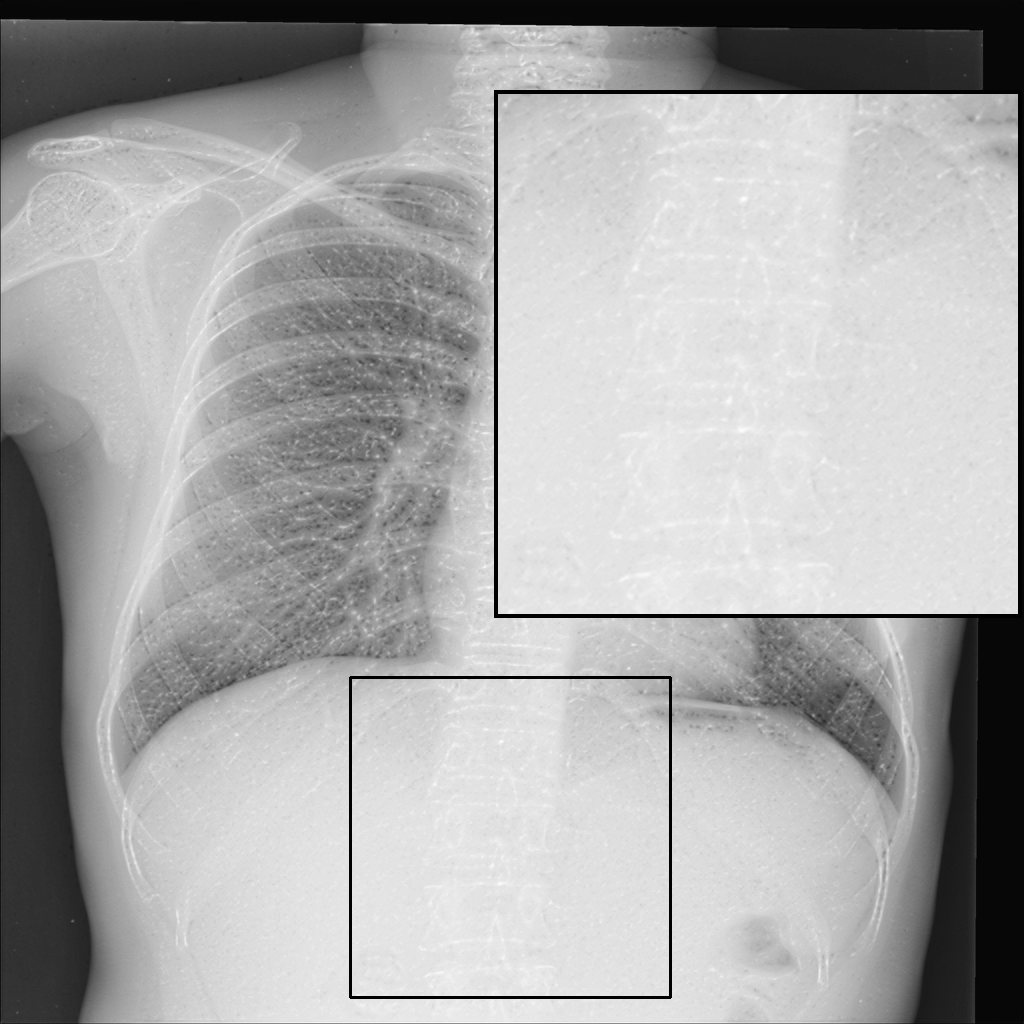} &
\includegraphics[width=\linewidth]{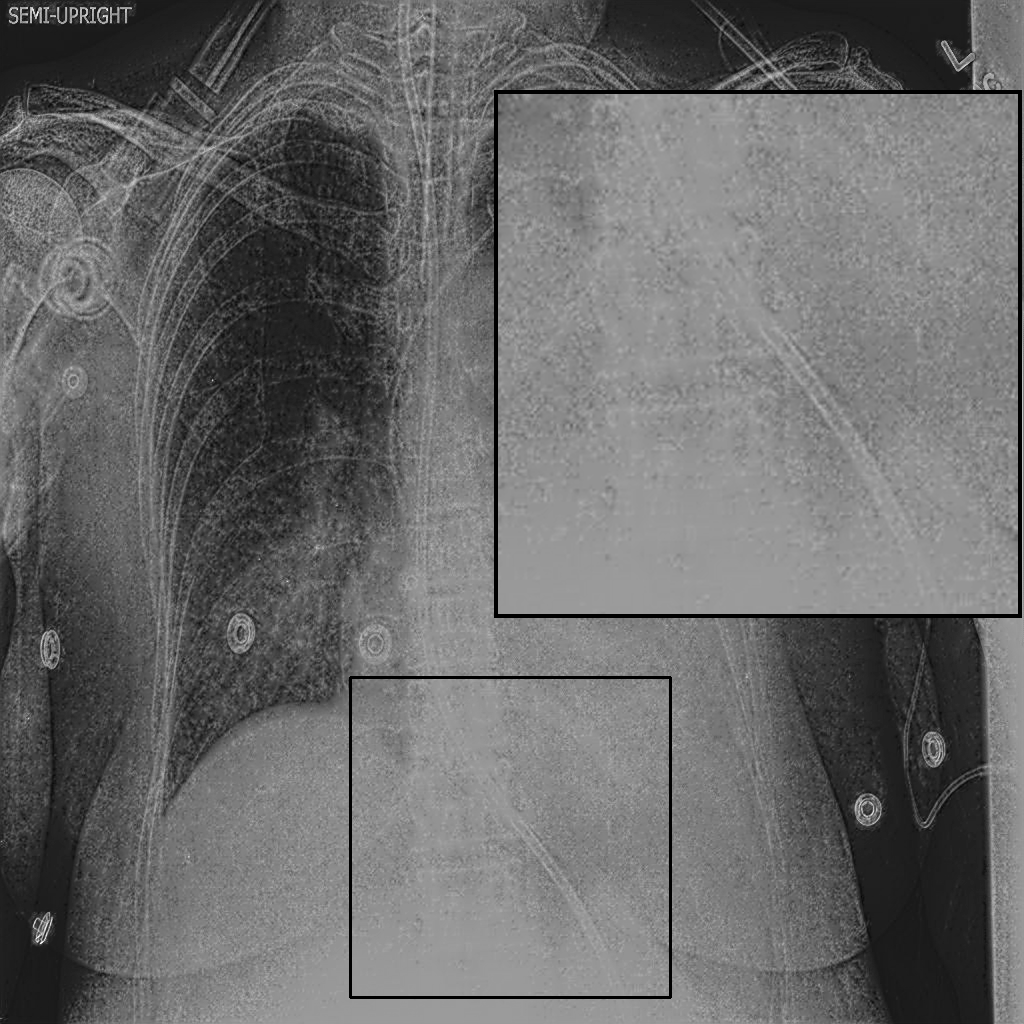}  \\

\multirow{2}{*}[0.07\columnwidth]{\rotatebox{90}{\textbf{XVertNet (Ours)}}} &
\includegraphics[width=\linewidth]{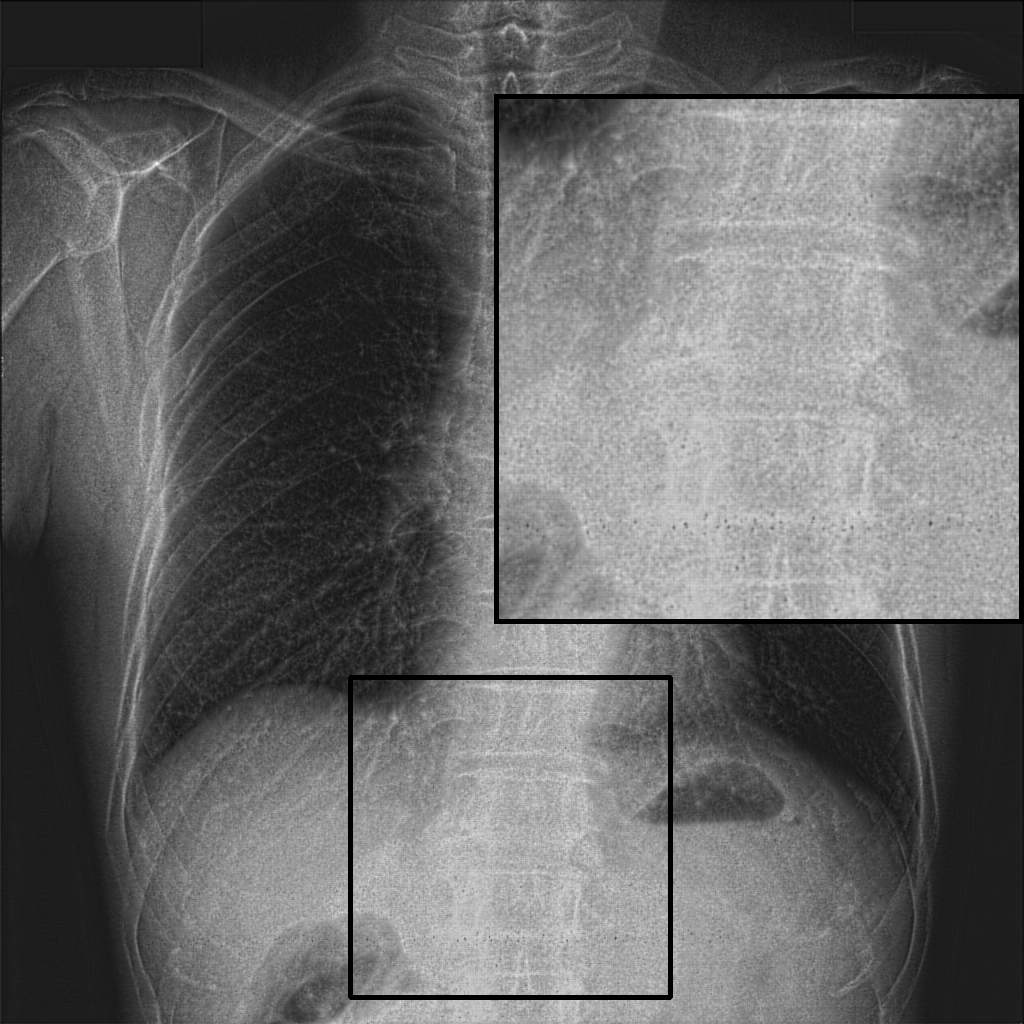} &
\includegraphics[width=\linewidth]{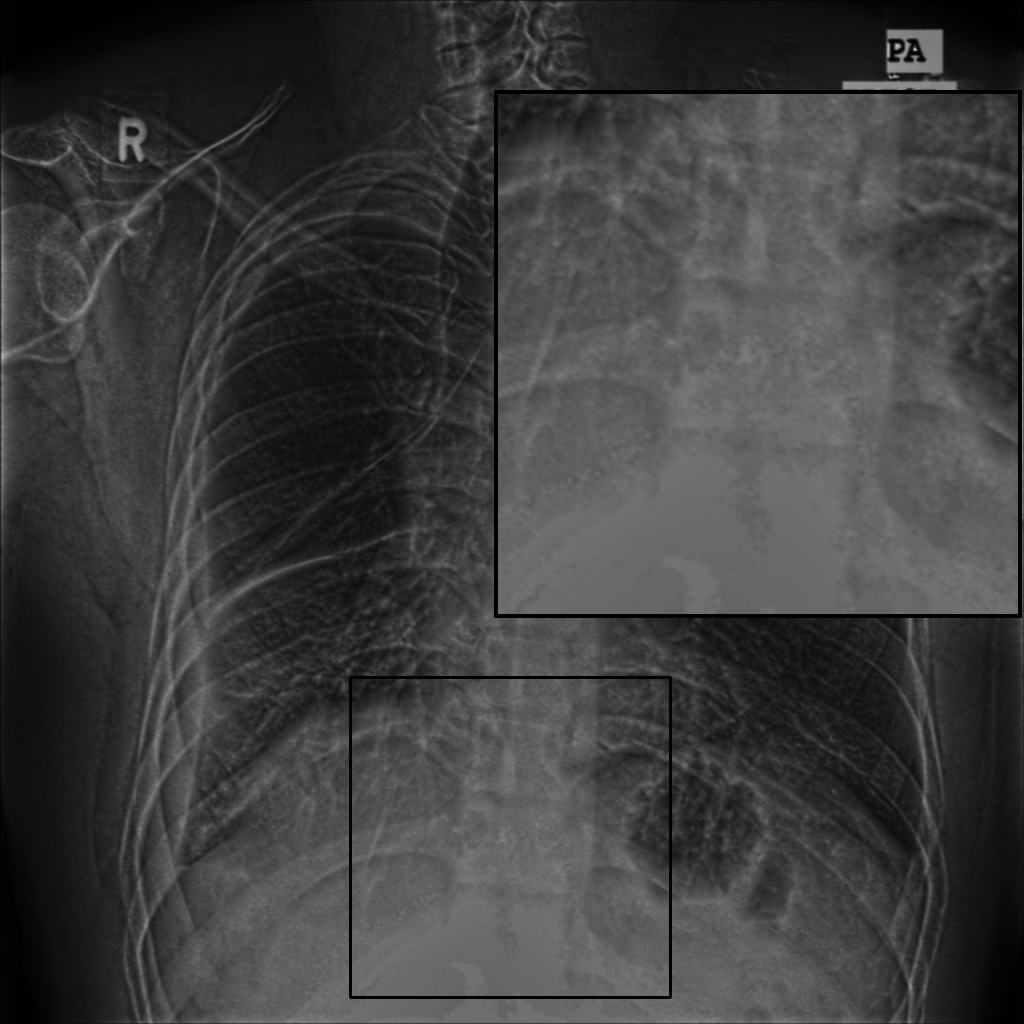} &
\includegraphics[width=\linewidth]{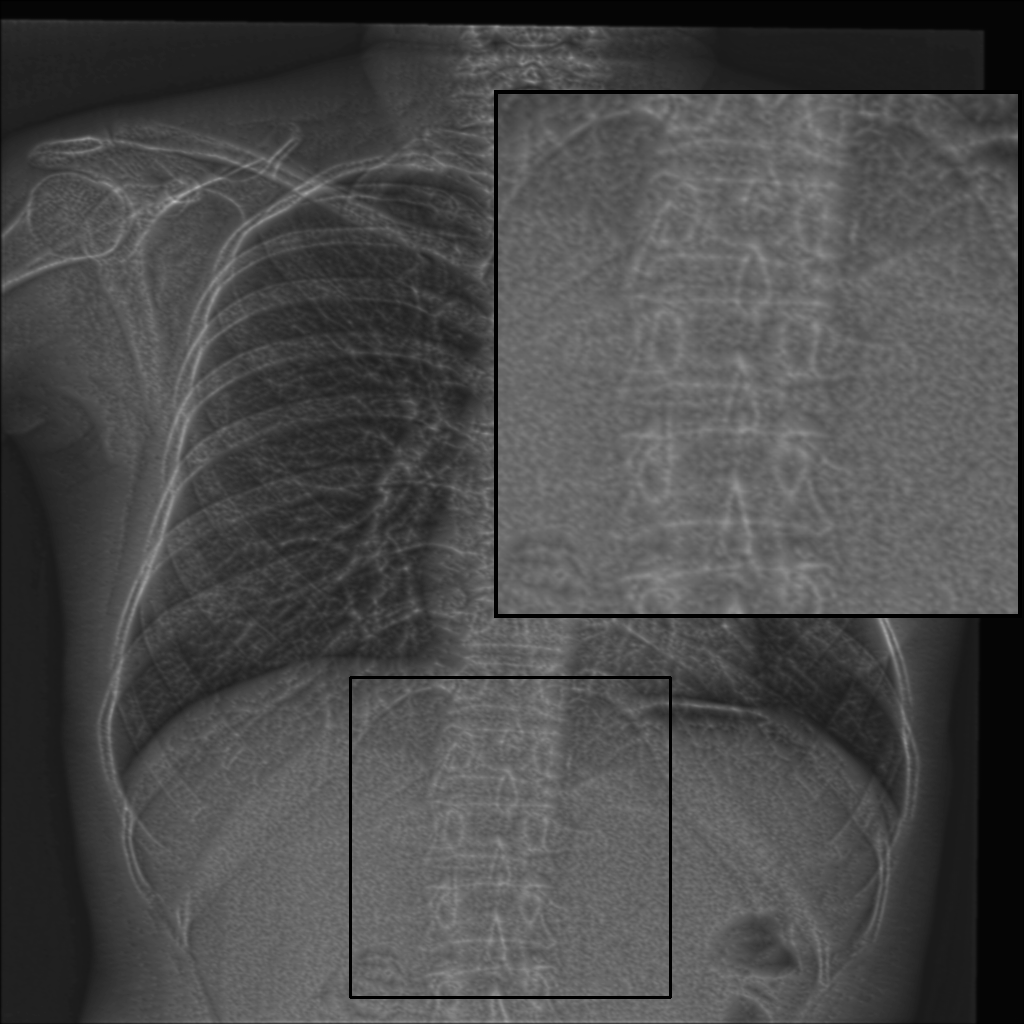} &
\includegraphics[width=\linewidth]{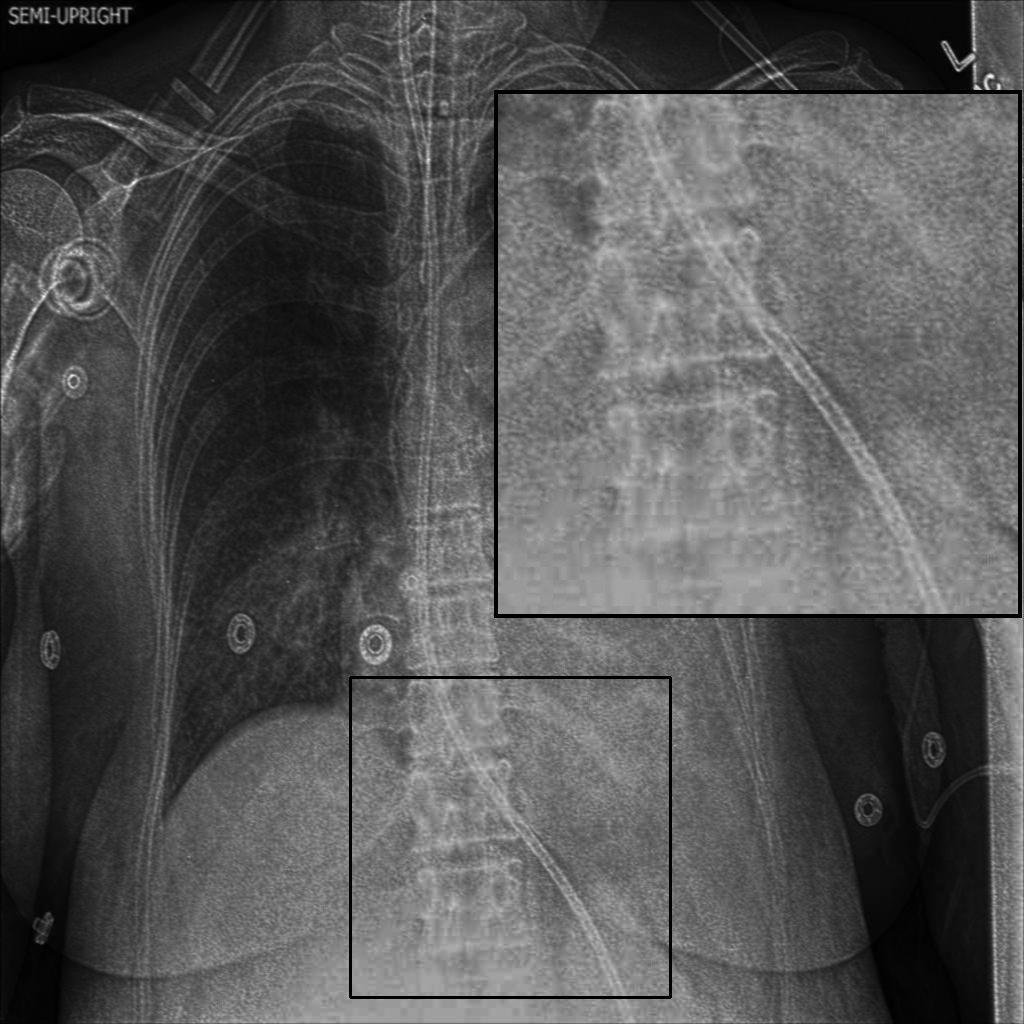}  \\
\end{tabular}

\endgroup
\caption{Visual examples of different image enhancement methods across datasets; first and second columns present results from the JSRT and Montgomery County Chest X-ray datasets, respectively, while third and fourth columns show results from the NIH Chest X-ray and CheXpert datasets}
\label{fig8}
\end{figure*}

\begin{table*}[ht!]
\centering
\caption{Comparison of performance between our model and traditional and learning-based methods on several benchmark datasets, expressed as mean $\pm$ std. deviation of various measures; bold and underlined values indicate best and second-best performance, respectively; $\ast$ indicates statistically significant difference ($p<0.05$) between XVertNet and another method for a given evaluation measure}

\label{table:1}

\resizebox{\textwidth}{!}{
\renewcommand{\arraystretch}{0.8}
\setlength{\tabcolsep}{3pt} 
{
\begin{tabular}{|c||c|c|c|c|c|c|c|c|}
\hline\hline\tiny
\tiny{Dataset} & \tiny{Measure} & \tiny{CLAHE \cite{Pisano1998}} & \tiny{Farbman \textit{et al.} \cite{Fattal2008}} & \tiny{ZSSR \cite{Shocher2018}} & \tiny{ Zero-DCE ~\cite{guo2020}} & \tiny{Madmad \textit{et al.}~\cite{Madmad2021}}  &\textbf{\tiny{XVertNet}} \\

\hline \hline
\multirow{4}{*}{\tiny{JSRT}} 

& \tiny{Entropy~$\uparrow$} & \tiny{6.501 $\pm$ 0.490*} & \tiny{4.891 $\pm$ 0.385*} & \tiny{5.587 $\pm$ 0.108*} &  \tiny{6.324 $\pm$ 0.489*} &\tiny{\underline{6.869 $\pm$ 0.400*}}  &
\tiny{\textbf{6.964 $\pm$ 0.158}} \\ \cline{2-8}

& \tiny{TC~$\uparrow$} & \tiny{0.020$\pm$ 0.008* } & \tiny{\textbf{0.170 $\pm$ 0.092}} & \tiny{0.160 $\pm$ 0.031} &  \tiny{0.015 $\pm$ 0.010*} & \tiny{0.154 $\pm$ 0.051} &
\tiny{\underline{0.161 $\pm$ 0.060}}\\ 
\cline{2-8}

& \tiny{LPC-SI~$\uparrow$} & \tiny{0.773 $\pm$ 0.034*} & \tiny{0.822 $\pm$ 0.067} & \tiny{0.654 $\pm$ 0.312*}  & \tiny{0.652 $\pm$ 0.119*} & \tiny{\underline{0.845 $\pm$ 0.117}} & \tiny{\textbf{0.868 $\pm$ 0.089}} \\ \cline{2-8}

& \tiny{TMQI~$\uparrow$} & \tiny{0.056 $\pm$ 0.032*} & \tiny{0.208 $\pm$ 0.002*} & \tiny{0.203 $\pm$ 0.002*}  & \tiny{\underline{0.211$\pm$ 0.001}} & \tiny{0.122 $\pm$ 0.001*} & \tiny{\textbf{0.212 $\pm$ 0.002}} \\ \cline{2-8}

& \tiny{{PIQE}~$\downarrow$} & \tiny{13.418 $\pm$ 1.771*} & \tiny{17.194 $\pm$ 2.457*} & \tiny{22.445 $\pm$ 5.736*} &  \tiny{\underline{7.217 $\pm$ 1.315}} & \tiny{13.428 $\pm$ 1.842*} & \tiny{\textbf{6.978 $\pm$ 1.260}} \\ \cline{2-8}
\hline

\multirow{4}{*}{\tiny{Montgomery}} 
& \tiny{Entropy~$\uparrow$} & \tiny{\underline{6.386 $\pm$ 0.749*}} & \tiny{4.402 $\pm$ 0.482*} & \tiny{3.883 $\pm$ 0.500*}   & \tiny{6.094 $\pm$ 0.942*} & \tiny{4.620 $\pm$ 0.225*} & \tiny{\textbf{7.385 $\pm$ 0.167}} \\ \cline{2-8}

& \tiny{TC~$\uparrow$} & \tiny{0.017 $\pm$ 0.006*} & \tiny{\textbf{0.062 $\pm$ 0.009}} & \tiny{0.038 $\pm$ 0.022*} & \tiny{0.014 $\pm$ 0.011} & \tiny{0.050 $\pm$ 0.002} & \tiny{\underline{0.059 $\pm$ 0.023}} \\\cline{2-8}

& \tiny{LPC-SI~$\uparrow$} & \tiny{0.799 $\pm$ 0.034*} & \tiny{\underline{0.852 $\pm$ 0.139}} & \tiny{0.704 $\pm$ 0.099*}  & \tiny{0.620 $\pm$ 0.125*} & \tiny{0.706 $\pm$ 0.055*} & \tiny{\textbf{0.856 $\pm$ 0.041}} \\ \cline{2-8}

& \tiny{TMQI~$\uparrow$} & \tiny{0.039 $\pm$ 0.006*} & \tiny{\underline{0.072 $\pm$ 0.012}} & \tiny{0.012 $\pm$ 0.008*}  & \tiny{0.068 $\pm$ 0.009} & \tiny{0.011 $\pm$ 0.003*} & \tiny{\textbf{0.148 $\pm$ 0.075}} \\ \cline{2-8}

& \tiny{{PIQE}~$\downarrow$} & \tiny{\textbf{9.980 $\pm$ 3.331}} & \tiny{13.614 $\pm$ 5.834*} & \tiny{10.282 $\pm$ 1.217}  & \tiny{15.918 $\pm$ 1.589*} & \tiny{10.238 $\pm$ 3.606} & \tiny{\underline{10.017 $\pm$ 2.998}} \\ \cline{2-8}
\hline

\multirow{4}{*}{\tiny{NIH}} 
& \tiny{Entropy~$\uparrow$} & \tiny{\underline{6.813 $\pm$ 0.465}} & \tiny{6.189 $\pm$ 0.714*} & \tiny{5.605 $\pm$ 0.696*}  & \tiny{6.262 $\pm$ 0.664*} & \tiny{6.591 $\pm$ 0.637} & \tiny{\textbf{6.813 $\pm$ 0.589}} \\ \cline{2-8}

& \tiny{TC~$\uparrow$} & \tiny{\underline{0.044 $\pm$ 0.015}} & \tiny{\textbf{0.116 $\pm$ 0.061*}} & \tiny{0.030 $\pm$ 0.021}  & \tiny{0.022 $\pm$ 0.011*} & \tiny{0.031 $\pm$ 0.022} & \tiny{0.037 $\pm$ 0.012} \\ \cline{2-8}

& \tiny{LPC-SI~$\uparrow$} & \tiny{\textbf{0.918 $\pm$ 0.029*}} & \tiny{\underline{0.910 $\pm$ 0.027}} & \tiny{0.883 $\pm$ 0.102*} & \tiny{0.790 $\pm$ 0.097*}  & \tiny{0.902 $\pm$ 0.022} & \tiny{0.908 $\pm$ 0.129} \\ \cline{2-8}

& \tiny{TMQI~$\uparrow$} & \tiny{0.074 $\pm$ 0.046*} & \tiny{\underline{0.201 $\pm$ 0.035}} & \tiny{0.111 $\pm$ 0.040*}  & \tiny{0.202 $\pm$ 0.020} & \tiny{0.200 $\pm$0.024} & \tiny{\textbf{0.207 $\pm$ 0.028}} \\ \cline{2-8}

& \tiny{{PIQE}~$\downarrow$} & \tiny{17.740 $\pm$ 4.455*} & \tiny{23.617 $\pm$ 7.609*} & \tiny{14.689 $\pm$ 1.105*} & \tiny{\underline{8.166 $\pm$ 2.697*}}  & \tiny{17.669 $\pm$ 4.133*} & \tiny{\textbf{7.076 $\pm$ 1.603}} \\ \cline{2-8}
\hline

\multirow{4}{*}{\tiny{CheXpert}} 
& \tiny{Entropy~$\uparrow$} & \tiny{\textbf{7.506 $\pm$ 0.261*}} & \tiny{6.594 $\pm$ 0.613*} & \tiny{6.517 $\pm$ 0.402*}  & \tiny{\underline{7.130 $\pm$ 0.338*}} & \tiny{6.516 $\pm$ 0.454*} & \tiny{6.973 $\pm$ 0.506} \\ \cline{2-8}

& \tiny{TC~$\uparrow$} & \tiny{0.108 $\pm$ 0.047*} & \tiny{\underline{0.219 $\pm$ 0.068}} & \tiny{0.038 $\pm$ 0.005*}  & \tiny{0.052$\pm$ 0.037*} & \tiny{0.115 $\pm$ 0.036*} & \tiny{ \textbf{0.222 $\pm$ 0.041}}\\  \cline{2-8}

& \tiny{LPC-SI~$\uparrow$} & \tiny{0.948 $\pm$ 0.012} & \tiny{\underline{0.950 $\pm$ 0.223*}} & \tiny{0.880 $\pm$ 0.023*}  & \tiny{0.909 $\pm$ 0.035*} & \tiny{0.889 $\pm$ 0.119*} & \tiny{\textbf{0.955 $\pm$ 0.034}} \\ \cline{2-8}

& \tiny{TMQI~$\uparrow$} & \tiny{0.026 $\pm$ 0.010*} & \tiny{0.139 $\pm$ 0.098*} & \tiny{0.094 $\pm$ 0.007*}  & \tiny{\underline{0.189 $\pm$ 0.069*}} & \tiny{0.152 $\pm$ 0.087*} & \tiny{\textbf{0.206 $\pm$ 0.027}} \\ \cline{2-8}

& \tiny{{PIQE}~$\downarrow$} & \tiny{23.506 $\pm$ 1.985} & \tiny{34.507 $\pm$ 6.481*} & \tiny{\textbf{9.193 $\pm$ 1.593*}}  & \tiny{\underline{14.053 $\pm$ 2.054*}} & \tiny{20.410 $\pm$ 7.154} & \tiny{22.958 $\pm$ 8.545} \\ \cline{2-8}
\hline \hline
\end{tabular}
}
}
\end{table*}

\begin{table*}[ht!]
\centering
\caption {Clinical experts cumulative ratings of image enhancement methods; bold indicates best performance in each column}

\label{table:2}

\resizebox{\textwidth}{!}{
\renewcommand{\arraystretch}{0.5}
\setlength{\tabcolsep}{3pt}
{
\begin{tabular}{|c||c|c|c||c|c|c|}
\hline\hline\tiny
& \multicolumn{3}{c||}{\tiny{Radiologist 1}} & \multicolumn{3}{c|}{\tiny{Radiologist 2}}\\

\hline\hline\tiny
\tiny{Scores} & \tiny{Farbman \textit{et al.} \cite{Fattal2008}} & \tiny{Madmad \textit{et al.}\cite{Madmad2021}}
& \tiny{\textbf{XVertNet}} & \tiny{Farbman \textit{et al.} \cite{Fattal2008}} & \tiny{Madmad \textit{et al.}\cite{Madmad2021}}
& \tiny{\textbf{XVertNet}} \\
\hline
\tiny{1} & \tiny{\textbf{73.91\%}} & \tiny{8.69\%} & \tiny{17.39\%} & \tiny{\textbf{43.47\%}} & \tiny{21.73\%} & \tiny{34.78\%} \\ 
\hline
\tiny{2} & \tiny{17.39\%} & \tiny{\textbf{56.52\%}} & \tiny{26.08\%} & \tiny{26.08\%} & \tiny{\textbf{52.17\%}} & \tiny{21.73\%}\\
\hline
\tiny{3} & \tiny{8.69\%} & \tiny{34.78\%} & \tiny{\textbf{56.52\%}} & \tiny{30.43\%} & \tiny{26.08\%} & \tiny{\textbf{43.47\%}} \\
\hline\hline
\end{tabular}

}
}
\end{table*}

We trained our model with 20,000 randomly selected images from the NIH ChestX-ray14 and CheXpert datasets. For validation, we used an additional 2,000 randomly selected images and evaluated the model's performance on 1,000 randomly sampled images from each of the two datasets. For the smaller JSRT and Montgomery County X-ray datasets, we split the data into 70\% for training, 20\% for validation, and 10\% for testing. The batch size was set to 8 images.

The Adam optimizer was used with a learning rate of 0.001, setting the parameters $\beta_{1}$ and $\beta_{2}$ to 0.9 and 0.99, respectively. Each training step consisted of 400 epochs for the large datasets (NIH ChestX-ray14 and CheXpert) and 50 epochs for the small datasets (JSRT and Montgomery County X-ray). Early stopping was applied when the validation loss did not improve over consecutive epochs to avoid overfitting and unnecessary training.

\subsection{Evaluation Criteria} 
\noindent In this study, we utilized entropy~\cite{Liu2014}, \textit{Tenengrad's criterion} (TC)~\cite{Buerkle2001}, the \textit{local phase coherence sharpness index} (LPC-SI)~\cite{Hassen2013},  \textit{tone-mapped image quality index} (TMQI)~\cite{Yeganeh2013},  and the \textit{perception-based image quality evaluator} (PIQE)~\cite{venkatanath2015}
to assess the strengths and capabilities of our proposed method.

As indicated previously, entropy is a statistical measure that serves as a reliable indicator of the information content present in an image. A high entropy value suggests that the image contains extensive texture and detail, as the gray levels are widely distributed over an entire range of values. Conversely, a low entropy value indicates that the gray levels are concentrated within a narrow range, implying relative uniformity and a lack of detail in the image. Typically, the entropy of an image is derived through an analysis of the probability distribution of its pixels.

The TC is based on gradient magnitude maximization and is considered one of the most robust and accurate image quality measures. The Tenengrad value of an image is calculated from the gradients at each pixel, where the partial derivatives are obtained using a high-pass filter, such as the Sobel operator. Typically, larger Tenengrad values indicate higher image quality. The TC has been commonly used to measure whether an image enhancement operator effectively improves the structural information of the image.

To evaluate the sharpness of an image, we used the LPC-SI to assess the coherence among local phase values in the image frequency domain at different spatial locations. Sharper images tend to exhibit higher local phase coherence because their frequency components are more consistent across space. An algorithm based on the LPC-SI involves passing the image through a series of log-Gabor filters, which detect the frequency content of the image at different scales and orientations. Based on the calculated local phase coherence values, the algorithm computes a sharpness index for the entire image, representing an overall measure of sharpness.

The TMQI combines structural fidelity with statistical naturalness to assess tone-mapped images. Structural fidelity evaluates how well a tone-mapping operator preserves important details by considering factors such as sharpness, texture preservation, and object arrangement. Statistical naturalness assesses how realistic the results appear in terms of brightness distribution and contrast. In general, higher TMQI values indicate better perceptual quality.

The PIQE measure assesses image perceptual quality by first computing mean-subtracted contrast-normalized
(MSCN) coefficients and analyzing only spatially active (i.e.,
high variance) blocks to mimic human visual focus. It detects
distortions and noise at the block level, assigns distortion
scores, and aggregates them to produce a no-reference quality score ranging from 0 to 100, where lower values indicate
better quality.

\subsection{Comparative Quantitative Performance}

We present in Table~\ref{table:1} a comparative quantitative performance evaluation. Specifically, we compared the performance of our proposed method with that of commonly used techniques, including CLAHE \cite{Pisano1998}, Farbman \textit{et al.}~\cite{Fattal2008}, ZSSR~\cite{Shocher2018}, Zero-DCE~\cite{guo2020}, and Madmad \textit{et al.} \cite{Madmad2021} across the four datasets described in Subsection~\ref{experimentsanalysis1}, namely: JSRT \cite{Shiraishi2000}, Montgomery County Chest X-ray~\cite{Jaeger2014}, NIH ChestX-ray14~\cite{wang2017}, and CheXpert \cite{Irvin2019}.

We employed the traditional non-learning methods of CLAHE~\cite{Pisano1998} and Farbman \textit{et al.}~\cite{Fattal2008} to full images. The learning-based method ZSSR \cite{Shocher2018} and Zero-DCE~\cite{guo2020} were trained on the same $512 \times 512$ cropped regions used for training XVertNet, while the method by Madmad \textit{et al.} \cite{Madmad2021} was trained on their original synthetic dataset. To ensure a fair comparative evaluation, we calculated the various evaluation measures on the same two $512 \times 290$ ROIs, corresponding to the upper and lower portions of the spinal cord, in the output images produced by each method.

For robust statistical evaluation, we applied $z$-score-based outlier removal, excluding values outside the range of mean $\pm$ std {$1.96$} (corresponding to a 95\% confidence interval). Outliers were identified separately for the enhanced and baseline methods, and the union of these sets was removed to ensure consistent comparisons. Statistical significance testing was then performed using the \textit{Wilcoxon ranked test}~\cite{Whitley2002}.

The quantitative results demonstrate that our proposed XVertNet consistently outperforms the baseline methods across most evaluation criteria. Specifically, XVertNet achieved the best performance in terms of entropy, LPC-SI, and TMQI, indicating superior information retention, enhanced structural sharpness, and improved perceptual quality, respectively. For the TC measure, XVertNet obtained the second-best performance after Farbman \textit{et al.}~\cite{Fattal2008}, who achieved the highest TC values across most datasets.

However, although the latter demonstrated higher TC scores, their method focuses primarily on gradient enhancement, often at the expense of overall image quality. This is reflected in their relatively lower entropy and TMQI values compared to XVertNet. In contrast, XVertNet balances local structure enhancement with global perceptual quality, producing more diagnostically reliable outputs.

The Wilcoxon rank test~\cite{Whitley2002} revealed statistically significant improvements ($p<0.05$) for XVertNet compared to CLAHE~\cite{Pisano1998}, ZSSR~\cite{Shocher2018}, Zero-DCE ~\cite{guo2020}, and Madmad \textit{et al.}~\cite{Madmad2021} across nearly all evaluation measures. Although Farbman \textit{et al.}~\cite{Fattal2008} achieved marginally better TC and LPC-SI scores in isolated cases, these differences were not statistically significant, highlighting the robustness and clinical relevance of our method in enhancing vertebral structures in chest X-rays.

\subsection{A Visual Assessment}\label{visualassessment}

Fig.~\ref{fig8} presents representative visual examples demonstrating the superiority of our method in enhancing vertebral contrast compared to several traditional and learning-based image enhancement techniques. The first and second columns show results from the JSRT and Montgomery County Chest X-ray datasets, while the third and fourth columns display results from the NIH ChestX-ray14 and CheXpert datasets. Across all datasets, the highlighted regions of interest (ROIs) emphasize the spinal areas, where fine structures and subtle details are critical for diagnostic assessment.
Our method, XVertNet, consistently reveals clearer and more distinct spinal structures without introducing notable artifacts, preserving both global and local anatomical information. In contrast, traditional methods such as CLAHE~\cite{Pisano1998} and Farbman \textit{et al.}~\cite{Fattal2008} often fail to enhance these subtle features effectively, either producing over-smoothed results or excessively sharpening noise. Similarly, learning-based methods such as ZSSR~\cite{Shocher2018} and Madmad \textit{et al.}~\cite{Madmad2021} exhibit difficulties in generalizing across different datasets, leading to either insufficient enhancement or visible artifacts. Zero-DCE~\cite{guo2020}, although effective for natural images, struggles to adapt to the uniform and low-contrast nature of medical X-rays and was not able to supply clear enough contrast-enhanced vertebral structures.

These visual comparisons underline the clinical relevance of our approach. XVertNet successfully enhances diagnostically important structures, particularly in challenging anatomical regions like the spine, where competing methods either distort the features or fail to sufficiently reveal them. This highlights the value of integrating targeted enhancement strategies that are specifically tailored to the unique characteristics of medical imaging data.

\subsection{Expert Qualitative Results}
We engaged two independent and unbiased clinicians to further assess the effectiveness of our proposed enhancement model from a clinical perspective. The evaluation was conducted on three sets of enhanced X-ray scans and one set of non-enhanced scans, each containing 69 randomly selected X-ray images from the JSRT, NIH ChestX-ray14, and CheXpert datasets. Each set included images enhanced by our model, as well as those produced by the methods of Farbman \textit{et al.} \cite{Fattal2008} and Madmad \textit{et al.} \cite{Madmad2021}, along with an unaltered set of original images. The radiologists, blinded to the enhancement methods, were asked to rank each image on a scale from 1 (lowest quality) to 3 (highest quality). The distribution of scores across the methods by Farbman \textit{et al.} \cite{Fattal2008}, Madmad \textit{et al.} \cite{Madmad2021}, and our method is summarized in Table~\ref{table:2}. Both radiologists consistently rated our method the highest, reinforcing its effectiveness in a professional clinical setting.

\subsection{Ablation Analysis}
We also conducted an ablation study to evaluate the impact of various components of our proposed methodology. The experiments were again carried out on the four merged datasets, JSRT, Montgomery County Chest X-ray, NIH ChestX-ray14, and CheXpert. Ablation analysis provides insight into the significance of different elements, particularly the GL, the choice of loss functions, and the number of learning steps. The superiority of our method is further demonstrated through performance comparisons in its different settings and visual assessments by expert clinicians.

\subsubsection{GL Effect} 
To evaluate the GL's impact, we conducted an ablation study by comparing model outputs with and without the GL after two training sessions (see Table~\ref{table:3}). The results, reported as mean $\pm$ standard deviation, demonstrate that the GL significantly improves image sharpness and perceptual quality, as reflected by higher TC, LPC-SI, and TMQI scores. The PIQE score also shows a dramatic decrease when the GL is included, indicating a reduction in perceptual image distortion. In contrast, entropy values remain comparable, suggesting a similar pixel distribution regardless of the presence of the GL.

\begin{figure}[ht!]
    \centering
    \begin{minipage}[b]{0.22\textwidth}
        \centering
        \includegraphics[width=\linewidth]{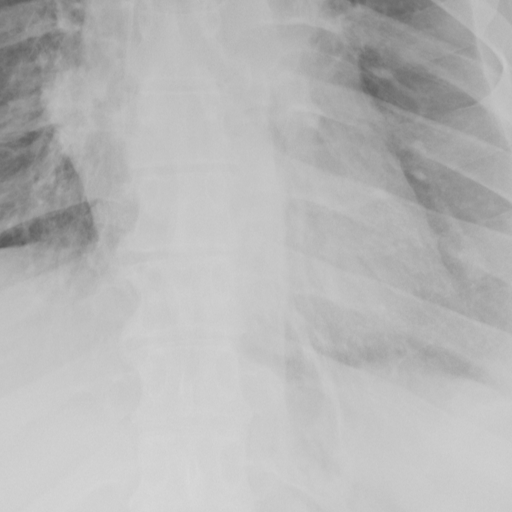}
        \caption*{\large{(a)}}
    \end{minipage}
    \hspace{0.01\textwidth} 
    \begin{minipage}[b]{0.22\textwidth}
        \centering
        \includegraphics[width=\linewidth]{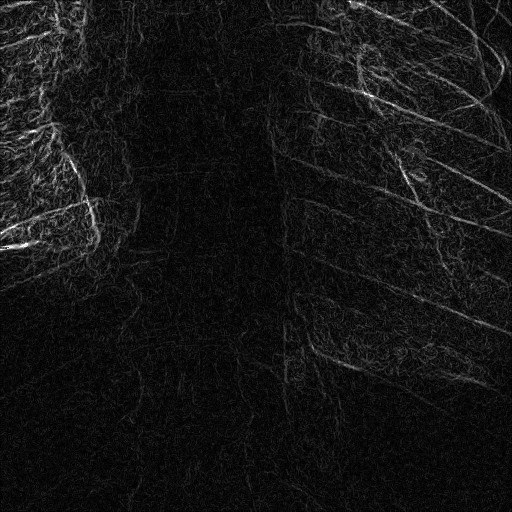}
        \caption*{\large{(b)}}
    \end{minipage}
    \hspace{0.01\textwidth}
    \begin{minipage}[b]{0.22\textwidth}
        \centering
        \includegraphics[width=\linewidth]{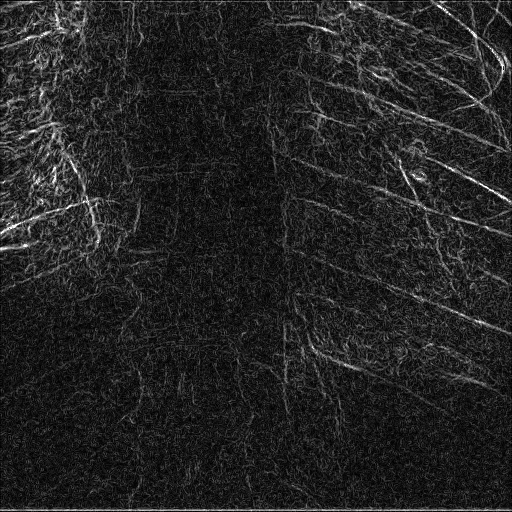}
        \caption*{\large{(c)}}
    \end{minipage}\\
    
    \hspace{0.01\textwidth}
    \begin{minipage}[b]{0.22\textwidth}
        \centering
        \includegraphics[width=\linewidth]{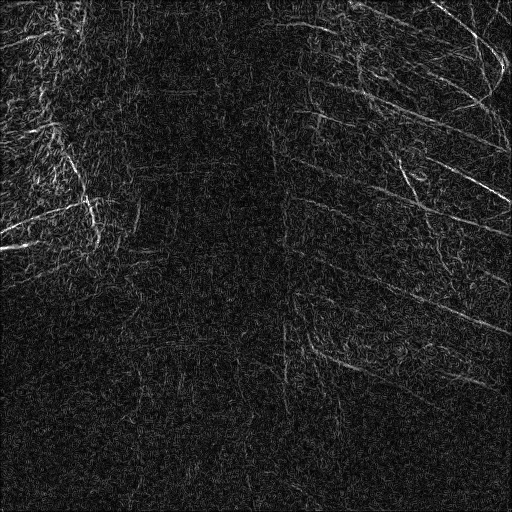}
        \caption*{\large{(d)}}
    \end{minipage}
    \hspace{0.01\textwidth}
    \begin{minipage}[b]{0.22\textwidth}
        \centering
        \includegraphics[width=\linewidth]{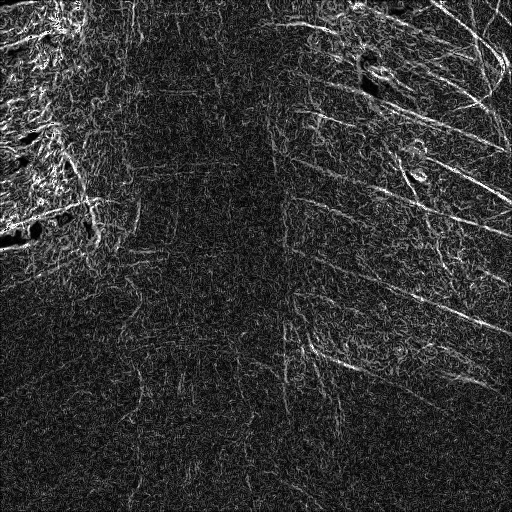}
        \caption*{\large{(e)}}
    \end{minipage}

    \caption{Visualization of gradient maps: (a) Original image fragment, (b) gradient of the original fragment, (c) gradient with single HFC as guidance, (d) gradient without GL, and (e) gradient with GL}
    \label{fig5}
\end{figure}

Statistical analysis using the Wilcoxon rank test (applied to scores within the range of mean $\pm$ 1.96 standard deviations) confirms the significance of improvements in TC, LPC-SI, TMQI, and PIQE ($p<0.05$). Interestingly, the model without GL exhibits significantly higher entropy ($p<0.05$), although this metric alone does not necessarily indicate superior structural or perceptual quality.  Furthermore, gradient map visualizations (see Fig.~\ref{fig5}) provide additional insight into the GL's contribution, revealing clearer object outlines and finer details in comparison to settings without the GL.

Overall, the ablation study underscores the substantial role of the GL in enhancing image quality, particularly in terms of TC, LPC-SI, and TMQI, as supported by both quantitative measures and qualitative visual evidence.

 \vspace{-0.01cm}
\begin{table}[ht!]
\centering
\caption{Statistical results for model with and without GL, expressed as mean $\pm$ std. deviation over various measures; bold indicates best performance in each column; $\ast$ indicates statistically significant difference ($p<0.05$) between the two models}

\label{table:3}
\setlength{\tabcolsep}{3pt}
\renewcommand{\arraystretch}{1.0}
\resizebox{0.7\textwidth}{!}{
 { 

\begin{tabular}{|c||c|c|c|c|c|c|}
\hline \hline \scriptsize

  \scriptsize{Mode}  & \scriptsize{Entropy~$\uparrow$} & \scriptsize{TC~$\uparrow$} &  \scriptsize{LPC-SI~$\uparrow$} & \scriptsize{TMQI~$\uparrow$} & \scriptsize{{PIQE}~$\downarrow$}\\
  
\hline \hline \scriptsize
\scriptsize{w/o GL}  
& \scriptsize{\textbf{7.224 $\pm$ 0.291*}}
& \scriptsize{0.056 $\pm$ 0.020*}
& \scriptsize{0.877 $\pm$ 0.167*} 
& \scriptsize{0.068 $\pm$ 0.010*} 
& \scriptsize{18.864 $\pm$ 1.181*}\\

\hline \scriptsize
\scriptsize{with GL}  
  & \scriptsize{7.082 $\pm$ 0.721}  
  & \scriptsize{\textbf{0.081 $\pm$ 0.047}} 
  & \scriptsize{\textbf{0.923 $\pm$ 0.056}}
  & \scriptsize{\textbf{0.071 $\pm$ 0.005}}
  & \scriptsize{\textbf{7.174 $\pm$ 0.420}}\\
\hline \hline 
\end{tabular}
}
}
\end{table}

\subsubsection{Optimized Loss}
To understand the contribution of each component in our loss design, we performed an ablation study comparing the gradient correlation loss $\mathcal{L}_{1}$, the feature enhancement loss $\mathcal{L}_{2}$, and their combination ($\mathcal{L}_{1} + \mathcal{L}_{2}$). The results, summarized in Table~\ref{table:4}, demonstrate the individual and joint effectiveness of these components across multiple image quality metrics.

Across all metrics, the combined configuration consistently outperformed the isolated losses. Notably, while the combined loss $\mathcal{L}_1 + \mathcal{L}_2$ resulted in statistically significant improvements across most evaluation metrics compared to both $\mathcal{L}_1$ and $\mathcal{L}_2$, the differences in TMQI and entropy relative to $\mathcal{L}_2$ were not statistically significant. This suggests that $\mathcal{L}_2$ already contributes strongly to perceptual contrast and structural enhancement. Nonetheless, the consistent gains across all other measures indicate that $\mathcal{L}_1$ and $\mathcal{L}_2$ offer complementary strengths, and their integration leads to a more balanced enhancement of contrast, detail preservation, and visual quality. Statistical significance was assessed using the paired Wilcoxon signed-rank test ($p < 0.05$).

\begin{table}[ht!]
\centering
\caption{Model’s performance for different loss settings, expressed as mean $\pm$ std. deviation over various measures; bold indicates best performance in each column; $\ast$ indicates statistically significant differences ($p<0.05$) between combined loss and individual loss components  }

\setlength{\tabcolsep}{3pt}
\renewcommand{\arraystretch}{1.0}
\resizebox{0.7\textwidth}{!}{
\label{table:4}
 { 
\begin{tabular}{|c|c|c|c|c|c|c|c|}

\hline \hline
  \scriptsize{Mode} & \scriptsize{Entropy~$\uparrow$}  & \scriptsize{TC~$\uparrow$} & \scriptsize{LPC-SI~$\uparrow$}
    & \scriptsize{TMQI~$\uparrow$} & \scriptsize{{PIQE}~$\downarrow$}\\

\hline \scriptsize
\scriptsize{$\mathcal{L}_{1}$} 
& \scriptsize{7.166 $\pm$ 0.279*} 
 & \scriptsize{0.059 $\pm$ 0.032*}
 & \scriptsize{0.890 $\pm$ 0.067*}
& \scriptsize{0.161 $\pm$ 0.061*}
& \scriptsize{25.773 $\pm$ 7.178*}
\\

\hline \hline \scriptsize
\scriptsize{$\mathcal{L}_{2}$}   
& \scriptsize{ 7.187 $\pm$ 0.918}
& \scriptsize{0.080 $\pm$ 0.015*}
& \scriptsize{0.876 $\pm$ 0.021*}
& \scriptsize{0.211 $\pm$ 0.002} 
& \scriptsize{15.699 $\pm$ 6.877*}\\

\hline \scriptsize
\scriptsize{$\mathcal{L}_{1}$+$\mathcal{L}_{2}$}
& \scriptsize{\textbf{7.230 $\pm$ 0.604}}
& \scriptsize{\textbf{0.103 $\pm$ 0.059}}
& \scriptsize{\textbf{0.910 $\pm$ 0.076}}  
& \scriptsize{\textbf{0.212 $\pm$ 0.003}}
& \scriptsize{\textbf{10.537 $\pm$ 2.001}}
 \\

\hline \hline 
\end{tabular}
}
}
\end{table}

\subsubsection{Effect of Iterative Training}
We conducted an ablation study to evaluate the influence of iterative training on model performance. Table~\ref{table:5} summarizes the results across three training iterations. The second iteration consistently outperformed both the first and third in terms of structural and perceptual quality metrics. 

Statistical significance, assessed using the Wilcoxon rank test (on values within $\pm1.96$ standard deviations of the mean), confirmed that improvements in TC, LPC-SI, TMQI, and PIQE from the second iteration were significant ($p<0.05$) when compared to both the first and third iterations. No statistically significant differences were found for entropy across iterations, although the second step still showed a marginal increase over the others.

Additionally, radiologists evaluated 60 randomly selected outputs (Table~\ref{table:6}) and consistently preferred those generated during the second iteration, assigning higher perceptual scores on a scale of 1 to 3. This further validates the quantitative improvements observed.

In summary, two iterations of training provide an optimal balance between contrast refinement and structural preservation, leading to clearer, diagnostically meaningful enhancements of vertebral regions.

\begin{table}[ht!]
\centering
\caption{Comparison of performance across three iterations, expressed as mean $\pm$ std. deviation over various measures; bold indicates best performance in each column; $\ast$ indicates statistically significant differences ($p<0.05$) between the second iteration and both the first and third iterations }
\label{table:5}

\setlength{\tabcolsep}{3pt}
\renewcommand{\arraystretch}{1.2}
\resizebox{0.7\textwidth}{!}{

\begin{tabular}{|c||c|c|c|c|c|c|}
\hline \hline\scriptsize

  \scriptsize{Mode} & \scriptsize{Entropy~$\uparrow$}  & \scriptsize{TC~$\uparrow$} &\scriptsize{LPC-SI~$\uparrow$}    & \scriptsize{TMQI~$\uparrow$} & \scriptsize{{PIQE}~$\downarrow$}
   \\

\hline \hline \scriptsize	
\scriptsize{1st step} 
& \scriptsize{7.100 $\pm$ 1.167}
& \scriptsize{0.043 $\pm$ 0.017*}
& \scriptsize{0.891  $\pm$ 0.012*}
& \scriptsize{0.166 $\pm$ 0.058*}
& \scriptsize{17.002 $\pm$ 6.087*}\\

\hline \scriptsize	
\scriptsize{2nd step} 
& \scriptsize{\textbf{7.123 $\pm$ 0.313}}
& \scriptsize{\textbf{0.067 $\pm$ 0.050}}
& \scriptsize{\textbf{0.923 $\pm$ 0.040}}
& \scriptsize{\textbf{0.197 $\pm$ 0.058}}
& \scriptsize{\textbf{15.696 $\pm$ 7.798}}\\
 
\hline \scriptsize	
\scriptsize{3rd step} 
& \scriptsize{7.063 $\pm$ 0.443}
 & \scriptsize{0.063 $\pm$ 0.008*}
& \scriptsize{0.850 $\pm$ 0.012*}
& \scriptsize{0.082 $\pm$ 0.055*}
& \scriptsize{22.577 $\pm$ 11.161*}\\ 
 \hline \hline
\end{tabular}
}
\end{table}

\begin{table}[ht!]
\centering
\caption{ Clinical experts' evaluations of the number of training steps (on the cumulative rating scale); bold indicates best performance in each column  }
\label{table:6}

\setlength{\tabcolsep}{3pt}
\renewcommand{\arraystretch}{1}
\resizebox{0.5\textwidth}{!}{
\begin{tabular}{|c||c|c|c||c|c|c|}
\hline \hline
   \scriptsize{} & \multicolumn{3}{c||} {\scriptsize{Radiologist 1}} & \multicolumn{3}{c|}{\scriptsize{Radiologist 2}}\\
\hline \scriptsize
  \scriptsize{Score}  & \scriptsize{1st step}
  & \scriptsize{2nd step} & \scriptsize{3rd step}
  & \scriptsize{1st step}
  & \scriptsize{2nd step} & \scriptsize{3rd step}
  \\

\hline \hline  \scriptsize
\scriptsize{1} &  
\scriptsize{\textbf{43.33\%}}
& \scriptsize{48.33\%} 
&   \scriptsize{30.00\%} 
& \textbf{\scriptsize{53.33\%}}
& \scriptsize{41.67\%}
&   \scriptsize{\textbf{86.67\%}}
\\

\hline \scriptsize
\scriptsize{2} 
& \scriptsize{31.67\%} 
& \scriptsize{0.00\%} 
& \textbf{\scriptsize{26.67\%}}
& \scriptsize{38.33\%} 
& \scriptsize{3.33\%}
&  \scriptsize{6.67\%}
\\

\hline \scriptsize
\scriptsize{3} 
& \scriptsize{25.00\%} 
& \scriptsize{\textbf{51.67\%}}
& \scriptsize{43.33\%}   
& \scriptsize{8.33\%}
&  \scriptsize{\textbf{55.00\%}}
&  \scriptsize{6.67\%}
\\ 
\hline \hline  

\end{tabular}
}
\end{table}

\section{Discussion and Conclusions} \label{discussion}

We presented a novel mechanism tailored to address the challenges of vertebral contrast in chest X-rays. At the core of our method is the \textit{dynamically self-adaptive guidance layer (GL)}—a carefully designed module that enables much better structural enhancement based on local image characteristics, without any external or prior knowledge. 
Our \textit{iterative training paradigm} reflects a deeper architectural philosophy, i.e., that enhancing anatomical structures cannot be achieved in a single pass. Instead, we progressively reinforce high-frequency features across stages, allowing the model to compound its representational depth. This iterative refinement acts as a self-corrective mechanism, aligning the enhancement process with the underlying anatomical structures more effectively than one-shot approaches.
We also introduced a \textit{dual-objective loss} that combines gradient correlation with entropy-based regularization. This loss simultaneously preserves anatomical contrast and suppresses noise-induced artifacts—two competing priorities in medical image enhancement. This duality is essential for preserving diagnostic quality in X-ray imaging, especially under varying acquisition conditions.
In contrast, classical methods such as CLAHE~\cite{Pisano1998} and gradient-based filters (e.g., Farbman \textit{et al.}~\cite{Fattal2008}) often suffer from over-sharpening or fail to generalize across datasets. Likewise, learning-based methods like ZSSR ~\cite{Shocher2018}, Madmad \textit{et al.}~\cite{Madmad2021}, and Zero-DCE ~\cite{guo2020} demonstrate inconsistent performance due to dataset sensitivity or reliance on synthetic training data.
Crucially, \textit{radiologist evaluations confirm the clinical relevance} of our framework. XVertNet-enhanced images consistently received the highest rankings, highlighting not only perceptual quality but also diagnostic utility. Lastly, our ablation studies reinforce the claim that the GL module, the dual-objective loss, and the multi-stage design are not interchangeable components—they are the result of deliberate architectural design aimed at solving a real clinical problem. To summarize, our proposed XVertNet is a robust, fully unsupervised method for vertebral enhancement in chest X-rays. By leveraging a self-tuned guidance layer, iterative refinement, and a multi-phase pipeline, it overcomes the limitations of commonly used techniques in the medical imaging domain. The consistent gains across diverse datasets and quality metrics demonstrate the method's scalability and effectiveness.

\bibliographystyle{IEEEtran}
\bibliography{biblio}

\end{document}